\documentclass[aps,prd,twocolumn,amsfonts,amssymb,amsmath,showpacs,letterpaper,superscriptaddress,nofootinbib,altaffilletter]{revtex4-1}

\usepackage{hyperref}

\usepackage{color}
\usepackage{graphicx}
\usepackage{latexsym}
\usepackage{float}
\usepackage{amsmath}
\usepackage{amssymb}
\usepackage{multirow}
\usepackage{ifthen}
\usepackage{slashed}

\newcommand{\hrss}{h_{\textrm{rss}}}

\newcommand{\xp}{\textsc{X-Pipeline}}
\newcommand{\cwb}{\textsc{cWB}}

\newcommand{\makevisible}[1]{\textcolor{red}{#1}}
\newcommand{\switch}[1]{
  \ifthenelse{\equal{#1}{0}}{\renewcommand{\makevisible}[1]{}}{}}
\switch{1} 
\def\version$#1,v #2 #3${#2}

\renewcommand{\today}{\number\day\space\ifcase\month\or
  January\or February\or March\or April\or May\or June\or
  July\or August\or September\or October\or November\or December\fi
  \space\number\year}

\def\be{\begin{equation}}
\def\ee{\end{equation}}
\def\bi{\begin{itemize}} 
\def\ei{\end{itemize}}
\def\ben{\begin{enumerate}}
\def\een{\end{enumerate}}
\def\hrssu{Hz${}^{-1/2}$}
\def\hrss{h_\mathrm{rss}}

\def\ligodoc{{LIGO-P14}{00208}{}} 

\def\aj{Astron. J.}
\def\apj{Astrophys. J.}
\def\apjl{Astrophys. J. Lett.}
\def\apjs{Astrophys. J. Supp. Ser. }

\def\aap{Astron. Astrophys. }
\def\araa{Ann.\ Rev. Astron. Astroph. }

\def\mnras{Mon. Not. Roy. Astron. Soc. }

\def\prl{Phys. Rev. Lett.}
\def\prd{Phys. Rev. D.}

\def\cqg{Class. Quantum Grav.}
\def\memsai{Mem. Societa Astronomica Italiana}

\begin{document}

\title{A First Targeted Search for Gravitational-Wave Bursts from Core-Collapse Supernovae in Data of First-Generation Laser Interferometer Detectors}

\begin{abstract}
We present results from a search for gravitational-wave bursts
coincident with 
two core-collapse supernovae observed optically in
2007 and 2011. We employ data from the Laser Interferometer Gravitational-wave 
Observatory (LIGO), the Virgo gravitational-wave observatory, and the GEO\,600 gravitational-wave observatory. 
The targeted core-collapse supernovae were selected
on the basis of (1) proximity (within approximately 15\,Mpc), (2)
tightness of observational constraints on the time of core collapse
that defines the gravitational-wave search window, and (3) coincident
operation of at least two interferometers at the time of core collapse.
We find no plausible gravitational-wave candidates. 
We present the probability of detecting signals from both astrophysically 
well-motivated and more speculative gravitational-wave emission mechanisms 
as a function of distance from Earth, and discuss the implications 
for the detection of gravitational waves from core-collapse supernovae 
by the upgraded Advanced LIGO and Virgo detectors.
\end{abstract}


\author{
B.~P.~Abbott,$^{1}$  
R.~Abbott,$^{1}$  
T.~D.~Abbott,$^{2}$  
M.~R.~Abernathy,$^{1}$  
F.~Acernese,$^{3,4}$ 
K.~Ackley,$^{5}$  
C.~Adams,$^{6}$  
T.~Adams,$^{7}$ 
P.~Addesso,$^{8}$  
R.~X.~Adhikari,$^{1}$
% ...
\\
\medskip
(LIGO Scientific Collaboration and Virgo Collaboration) 
\\
\medskip
{{}$^{\dag}$Deceased, May 2015. {}$^{\ddag}$Deceased, March 2015. }
}\noaffiliation
\affiliation {LIGO, California Institute of Technology, Pasadena, CA 91125, USA }
\affiliation {Louisiana State University, Baton Rouge, LA 70803, USA }
\affiliation {Universit\`a di Salerno, Fisciano, I-84084 Salerno, Italy }
\affiliation {INFN, Sezione di Napoli, Complesso Universitario di Monte S.Angelo, I-80126 Napoli, Italy }
\affiliation {University of Florida, Gainesville, FL 32611, USA }
\affiliation {LIGO Livingston Observatory, Livingston, LA 70754, USA }
\affiliation {Laboratoire d'Annecy-le-Vieux de Physique des Particules (LAPP), Universit\'e Savoie Mont Blanc, CNRS/IN2P3, F-74941 Annecy-le-Vieux, France }
\affiliation {University of Sannio at Benevento, I-82100 Benevento, Italy and INFN, Sezione di Napoli, I-80100 Napoli, Italy }
\affiliation {Albert-Einstein-Institut, Max-Planck-Institut f\"ur Gravi\-ta\-tions\-physik, D-30167 Hannover, Germany }
\affiliation {Nikhef, Science Park, 1098 XG Amsterdam, The Netherlands }
\affiliation {LIGO, Massachusetts Institute of Technology, Cambridge, MA 02139, USA }
\affiliation {Instituto Nacional de Pesquisas Espaciais, 12227-010 S\~{a}o Jos\'{e} dos Campos,S\~{a}o Paulo, Brazil }
\affiliation {INFN, Gran Sasso Science Institute, I-67100 L'Aquila, Italy }
\affiliation {INFN, Sezione di Roma Tor Vergata, I-00133 Roma, Italy }
\affiliation {Inter-University Centre for Astronomy and Astrophysics, Pune 411007, India }
\affiliation {International Centre for Theoretical Sciences, Tata Institute of Fundamental Research, Bangalore 560012, India }
\affiliation {University of Wisconsin-Milwaukee, Milwaukee, WI 53201, USA }
\affiliation {Leibniz Universit\"at Hannover, D-30167 Hannover, Germany }
\affiliation {Universit\`a di Pisa, I-56127 Pisa, Italy }
\affiliation {INFN, Sezione di Pisa, I-56127 Pisa, Italy }
\affiliation {Australian National University, Canberra, Australian Capital Territory 0200, Australia }
\affiliation {The University of Mississippi, University, MS 38677, USA }
\affiliation {California State University Fullerton, Fullerton, CA 92831, USA }
\affiliation {LAL, Univ. Paris-Sud, CNRS/IN2P3, Universit\'e Paris-Saclay, Orsay, France }
\affiliation {Chennai Mathematical Institute, Chennai 603103, India }
\affiliation {Universit\`a di Roma Tor Vergata, I-00133 Roma, Italy }
\affiliation {University of Southampton, Southampton SO17 1BJ, United Kingdom }
\affiliation {Universit\"at Hamburg, D-22761 Hamburg, Germany }
\affiliation {INFN, Sezione di Roma, I-00185 Roma, Italy }
\affiliation {Albert-Einstein-Institut, Max-Planck-Institut f\"ur Gravitations\-physik, D-14476 Potsdam-Golm, Germany }
\affiliation {APC, AstroParticule et Cosmologie, Universit\'e Paris Diderot, CNRS/IN2P3, CEA/Irfu, Observatoire de Paris, Sorbonne Paris Cit\'e, F-75205 Paris Cedex 13, France }
\affiliation {Montana State University, Bozeman, MT 59717, USA }
\affiliation {Universit\`a di Perugia, I-06123 Perugia, Italy }
\affiliation {INFN, Sezione di Perugia, I-06123 Perugia, Italy }
\affiliation {European Gravitational Observatory (EGO), I-56021 Cascina, Pisa, Italy }
\affiliation {Syracuse University, Syracuse, NY 13244, USA }
\affiliation {SUPA, University of Glasgow, Glasgow G12 8QQ, United Kingdom }
\affiliation {LIGO Hanford Observatory, Richland, WA 99352, USA }
\affiliation {Wigner RCP, RMKI, H-1121 Budapest, Konkoly Thege Mikl\'os \'ut 29-33, Hungary }
\affiliation {Columbia University, New York, NY 10027, USA }
\affiliation {Stanford University, Stanford, CA 94305, USA }
\affiliation {Universit\`a di Padova, Dipartimento di Fisica e Astronomia, I-35131 Padova, Italy }
\affiliation {INFN, Sezione di Padova, I-35131 Padova, Italy }
\affiliation {CAMK-PAN, 00-716 Warsaw, Poland }
\affiliation {University of Birmingham, Birmingham B15 2TT, United Kingdom }
\affiliation {Universit\`a degli Studi di Genova, I-16146 Genova, Italy }
\affiliation {INFN, Sezione di Genova, I-16146 Genova, Italy }
\affiliation {RRCAT, Indore MP 452013, India }
\affiliation {Faculty of Physics, Lomonosov Moscow State University, Moscow 119991, Russia }
\affiliation {SUPA, University of the West of Scotland, Paisley PA1 2BE, United Kingdom }
\affiliation {University of Western Australia, Crawley, Western Australia 6009, Australia }
\affiliation {Department of Astrophysics/IMAPP, Radboud University Nijmegen, P.O. Box 9010, 6500 GL Nijmegen, The Netherlands }
\affiliation {Artemis, Universit\'e C\^ote d'Azur, CNRS, Observatoire C\^ote d'Azur, CS 34229, Nice cedex 4, France }
\affiliation {MTA E\"otv\"os University, ``Lendulet'' Astrophysics Research Group, Budapest 1117, Hungary }
\affiliation {Institut de Physique de Rennes, CNRS, Universit\'e de Rennes 1, F-35042 Rennes, France }
\affiliation {Washington State University, Pullman, WA 99164, USA }
\affiliation {Universit\`a degli Studi di Urbino 'Carlo Bo', I-61029 Urbino, Italy }
\affiliation {INFN, Sezione di Firenze, I-50019 Sesto Fiorentino, Firenze, Italy }
\affiliation {University of Oregon, Eugene, OR 97403, USA }
\affiliation {Laboratoire Kastler Brossel, UPMC-Sorbonne Universit\'es, CNRS, ENS-PSL Research University, Coll\`ege de France, F-75005 Paris, France }
\affiliation {Astronomical Observatory Warsaw University, 00-478 Warsaw, Poland }
\affiliation {VU University Amsterdam, 1081 HV Amsterdam, The Netherlands }
\affiliation {University of Maryland, College Park, MD 20742, USA }
\affiliation {Center for Relativistic Astrophysics and School of Physics, Georgia Institute of Technology, Atlanta, GA 30332, USA }
\affiliation {Laboratoire des Mat\'eriaux Avanc\'es (LMA), CNRS/IN2P3, F-69622 Villeurbanne, France }
\affiliation {Universit\'e Claude Bernard Lyon 1, F-69622 Villeurbanne, France }
\affiliation {Universitat de les Illes Balears, IAC3---IEEC, E-07122 Palma de Mallorca, Spain }
\affiliation {Universit\`a di Napoli 'Federico II', Complesso Universitario di Monte S.Angelo, I-80126 Napoli, Italy }
\affiliation {NASA/Goddard Space Flight Center, Greenbelt, MD 20771, USA }
\affiliation {Canadian Institute for Theoretical Astrophysics, University of Toronto, Toronto, Ontario M5S 3H8, Canada }
\affiliation {Tsinghua University, Beijing 100084, China }
\affiliation {University of Michigan, Ann Arbor, MI 48109, USA }
\affiliation {National Tsing Hua University, Hsinchu City, 30013 Taiwan, Republic of China }
\affiliation {Charles Sturt University, Wagga Wagga, New South Wales 2678, Australia }
\affiliation {University of Chicago, Chicago, IL 60637, USA }
\affiliation {Caltech CaRT, Pasadena, CA 91125, USA }
\affiliation {Korea Institute of Science and Technology Information, Daejeon 305-806, Korea }
\affiliation {Carleton College, Northfield, MN 55057, USA }
\affiliation {Universit\`a di Roma 'La Sapienza', I-00185 Roma, Italy }
\affiliation {University of Brussels, Brussels 1050, Belgium }
\affiliation {Sonoma State University, Rohnert Park, CA 94928, USA }
\affiliation {Texas Tech University, Lubbock, TX 79409, USA }
\affiliation {Cardiff University, Cardiff CF24 3AA, United Kingdom }
\affiliation {University of Minnesota, Minneapolis, MN 55455, USA }
\affiliation {The University of Melbourne, Parkville, Victoria 3010, Australia }
\affiliation {The University of Texas Rio Grande Valley, Brownsville, TX 78520, USA }
\affiliation {The University of Sheffield, Sheffield S10 2TN, United Kingdom }
\affiliation {Universit\`a di Trento, Dipartimento di Fisica, I-38123 Povo, Trento, Italy }
\affiliation {INFN, Trento Institute for Fundamental Physics and Applications, I-38123 Povo, Trento, Italy }
\affiliation {Montclair State University, Montclair, NJ 07043, USA }
\affiliation {The Pennsylvania State University, University Park, PA 16802, USA }
\affiliation {National Astronomical Observatory of Japan, 2-21-1 Osawa, Mitaka, Tokyo 181-8588, Japan }
\affiliation {School of Mathematics, University of Edinburgh, Edinburgh EH9 3FD, United Kingdom }
\affiliation {Indian Institute of Technology, Gandhinagar Ahmedabad Gujarat 382424, India }
\affiliation {Institute for Plasma Research, Bhat, Gandhinagar 382428, India }
\affiliation {University of Szeged, D\'om t\'er 9, Szeged 6720, Hungary }
\affiliation {Embry-Riddle Aeronautical University, Prescott, AZ 86301, USA }
\affiliation {Tata Institute of Fundamental Research, Mumbai 400005, India }
\affiliation {INAF, Osservatorio Astronomico di Capodimonte, I-80131, Napoli, Italy }
\affiliation {American University, Washington, D.C. 20016, USA }
\affiliation {University of Massachusetts-Amherst, Amherst, MA 01003, USA }
\affiliation {University of Adelaide, Adelaide, South Australia 5005, Australia }
\affiliation {West Virginia University, Morgantown, WV 26506, USA }
\affiliation {University of Bia{\l }ystok, 15-424 Bia{\l }ystok, Poland }
\affiliation {SUPA, University of Strathclyde, Glasgow G1 1XQ, United Kingdom }
\affiliation {IISER-TVM, CET Campus, Trivandrum Kerala 695016, India }
\affiliation {Northwestern University, Evanston, IL 60208, USA }
\affiliation {Institute of Applied Physics, Nizhny Novgorod, 603950, Russia }
\affiliation {Pusan National University, Busan 609-735, Korea }
\affiliation {Hanyang University, Seoul 133-791, Korea }
\affiliation {NCBJ, 05-400 \'Swierk-Otwock, Poland }
\affiliation {IM-PAN, 00-956 Warsaw, Poland }
\affiliation {Rochester Institute of Technology, Rochester, NY 14623, USA }
\affiliation {Monash University, Victoria 3800, Australia }
\affiliation {Seoul National University, Seoul 151-742, Korea }
\affiliation {University of Alabama in Huntsville, Huntsville, AL 35899, USA }
\affiliation {ESPCI, CNRS, F-75005 Paris, France }
\affiliation {Universit\`a di Camerino, Dipartimento di Fisica, I-62032 Camerino, Italy }
\affiliation {Southern University and A\&M College, Baton Rouge, LA 70813, USA }
\affiliation {College of William and Mary, Williamsburg, VA 23187, USA }
\affiliation {Instituto de F\'\i sica Te\'orica, University Estadual Paulista/ICTP South American Institute for Fundamental Research, S\~ao Paulo SP 01140-070, Brazil }
\affiliation {University of Cambridge, Cambridge CB2 1TN, United Kingdom }
\affiliation {IISER-Kolkata, Mohanpur, West Bengal 741252, India }
\affiliation {Rutherford Appleton Laboratory, HSIC, Chilton, Didcot, Oxon OX11 0QX, United Kingdom }
\affiliation {Whitman College, 345 Boyer Avenue, Walla Walla, WA 99362 USA }
\affiliation {National Institute for Mathematical Sciences, Daejeon 305-390, Korea }
\affiliation {Universit\'e de Lyon, F-69361 Lyon, France }
\affiliation {Hobart and William Smith Colleges, Geneva, NY 14456, USA }
\affiliation {Janusz Gil Institute of Astronomy, University of Zielona G\'ora, 65-265 Zielona G\'ora, Poland }
\affiliation {Andrews University, Berrien Springs, MI 49104, USA }
\affiliation {Universit\`a di Siena, I-53100 Siena, Italy }
\affiliation {Trinity University, San Antonio, TX 78212, USA }
\affiliation {University of Washington, Seattle, WA 98195, USA }
\affiliation {Kenyon College, Gambier, OH 43022, USA }
\affiliation {Abilene Christian University, Abilene, TX 79699, USA }

\date[\relax]{Dated: \today }

\maketitle

\section{Introduction}
\label{sec:introduction}

Core-collapse supernovae (CCSNe) mark the violent death of massive
stars. It is believed that the initial collapse of a star's iron core
results in the formation of a proto-neutron star and the launch of a
hydrodynamic shock wave.  The latter, however, fails to immediately
explode the star, but stalls and must be \emph{revived} by a 
yet-uncertain supernova ``mechanism'' on a $\sim$$0.5-1\,\mathrm{s}$
timescale to explode the star (e.g.,
\cite{bethe:90,janka:12a,burrows:13a}). If the shock is not revived, a
black hole is formed and no or only a very weak explosion results
(e.g., \cite{oconnor:11,lovegrove:13,piro:13}). If the shock is
revived, it reaches the stellar surface and produces the spectacular
electromagnetic display of a Type II or Type Ib/c supernova.  The Type
classification is based on the explosion light curve and spectrum, which depend
largely on the nature of the progenitor star (e.g.,
\cite{filippenko:97}). The time from core collapse to breakout of the
shock through the stellar surface and first supernova light is
minutes to days, depending on the radius of the progenitor and energy
of the explosion (e.g., \cite{kistler:13,matzner:99,morozova:15}).

Any core collapse event generates a burst of neutrinos that
releases most of the proto-neutron star's gravitational binding energy
($\sim 3 \times 10^{53}\,\mathrm{erg} \approx 0.15\,M_\odot c^2$) on a
timescale of order $10$~seconds. This neutrino burst was detected from
SN 1987A and confirmed the basic theory of CCSNe
\cite{baade:34a,bethe:90,hirata:87,bionta:87}.

Gravitational waves (GWs) are emitted by aspherical mass-energy
dynamics that includes quadrupole or higher-order contributions. Such
asymmetric dynamics are expected to be present in the pre-explosion
stalled-shock phase of CCSNe and may be crucial to the CCSN explosion
mechanism (see, e.g., \cite{bhf:95,herant:95,couch:15,lentz:15}). GWs
can serve as probes of the magnitude and character of these
asymmetries and thus may help in constraining the CCSN mechanism
\cite{ott:09b,logue:12,abdikamalov:14}. 

Stellar collapse and CCSNe were considered as potential sources of
detectable GWs already for resonant bar detectors in the 1960s
\cite{weber:66}. Early analytic and semi-analytic estimates of the GW
signature of stellar collapse and CCSNe (e.g.,
\cite{ruffini:71,saenzshapiro:79,thuan:74,ipsermanagan:84,thorne:87})
gave optimistic signal strengths, suggesting that first-generation
laser interferometer detectors could detect GWs from CCSNe in the
Virgo cluster (at distances $D\gtrsim 10\,\mathrm{Mpc}$). Modern
detailed multi-dimensional CCSN simulations (see, e.g.,
\cite{dimmelmeier:08,yakunin:10,ott:11a,mueller:e12,mueller:13gw,ott:12a,ott:13a,abdikamalov:14,kuroda:14,yakunin:15}
and the reviews in \cite{ott:09,kotake:13review,fryernew:11}) find GW
signals of short duration ($\lesssim 1\,\mathrm{s}$) and emission
frequencies in the most sensitive $\sim$$10-2000\,\mathrm{Hz}$ band of
ground based laser interferometer detectors. Predicted total emitted
GW energies are in the range $10^{-12}-10^{-8}\, M_\odot c^2$ for
emission mechanisms and progenitor parameters that are presently
deemed realistic. These numbers suggest that the early predictions were optimistic and that
even second-generation laser interferometers (operating from 2015$+$) such as Advanced LIGO \cite{aLIGO}, Advanced
Virgo \cite{aVirgo}, and KAGRA \cite{kagra} will only be able to detect GWs from very nearby CCSNe at $D \lesssim 1-100\,\mathrm{kpc}$. 
Only our own Milky Way and the Magellanic Clouds are within that
range. The expected event rate is very low and estimated to $\lesssim
2-3\,\mathrm{CCSNe}/100\,\mathrm{yr}$
\cite{vandenbergh:91,cappellaro:93,tammann:94,li:11,diehl:06,maoz:10}.

However, there are also a number of analytic and semi-analytic GW emission
models of more extreme scenarios, involving non-axisymmetric
rotational instabilities, centrifugal fragmentation, and accretion
disk instabilities. The emitted GW signals may be sufficiently strong
to be detectable to much greater distances of $D \gtrsim
10-15\,\mathrm{Mpc}$, perhaps even with first-generation laser
interferometers (e.g.,
\cite{ott:06prl,fryer:02,piro:07,vanputten:04}).  These emission
scenarios require special and rare progenitor characteristics, but
they cannot presently be strictly ruled out on theoretical grounds. In a sphere of radius $\sim 15\,\mathrm{Mpc}$ centered on
Earth, the CCSN rate is $\gtrsim 1/\mathrm{yr}$
\cite{kistler:13,ando:05}. This makes Virgo cluster CCSNe interesting
targets for constraining extreme GW emission scenarios.

Previous observational constraints on GW burst sources applicable to
CCSNe come from all-sky searches for short-duration GW burst
signals
\cite{ando:05tama,FirstBurst,abbott:07burst,ligo_hfburst:09,ligo_burst_s5y1:09,S5y2Burst,S6Burst}.
These searches did not target individual astrophysical events. 
Targeted searches have the advantage over all-sky searches that
potential signal candidates 
in the data streams have to
arrive in a well-defined temporal \emph{on-source window} and have to
be consistent with coming from the sky location of the source. Both
constraints can significantly reduce the noise background and
improve the sensitivity of the search (e.g., \cite{sutton:10}).
Previous targeted GW searches have been carried out for gamma-ray
bursts
\cite{abbott:08m31,abbott:08grb,abadie_s5sgrb:10,ligo_051103:12,abadie:12grb,aasi:13stampgrb,aasi:14grb,aasi:14ipngrb},
soft-gamma repeater flares \cite{abbott:08sgr,ligo_a5_sgr:11}, and
pulsar glitches \cite{velaglitch:11}. 
A recent study \cite{gossan:16} confirmed that targeted searches with 
Advanced LIGO and Virgo at design sensitivity should be able to detect 
neutrino-driven CCSNe out to several kiloparsecs 
and rapidly rotating CCSNe out to tens of kiloparsecs, 
while more extreme GW emission scenarios will be detectable to several megaparsecs.

In this paper, we present the first targeted search for GWs from CCSNe
using the first-generation Initial LIGO (iLIGO)~\cite{LIGO}, GEO\,600~\cite{grote:10},
and Virgo~\cite{VIRGO} laser interferometer detectors. The data
searched were collected over 2005--2011 in the S5, A5, and S6 runs of
the iLIGO and GEO\,600 detectors, and in the VSR1--VSR4 runs of the
Virgo detector. From the set of CCSNe observed in this period
\cite{snlistweb}, we make a preliminary selection of four targets for
our search: SNe~2007gr, 2008ax, 2008bk, and 2011dh. These CCSNe
exploded in nearby galaxies ($D\lesssim10\,\mathrm{Mpc}$), have well
constrained explosion dates, and at least partial coverage by
coincident observation of more than one interferometer. SNe 2008ax and
2008bk occurred in the \emph{astrowatch} (A5) period between the S5
and S6 iLIGO science runs. In A5, the principal goal was detector
commissioning, not data collection. Data quality and sensitivity were
not of primary concern.  Preliminary analyses of the
gravitational-wave data associated with SNe 2008ax and 2008bk showed
that the sensitivity was much poorer than the data for SNe~2007gr and
2011dh.  Because of this, we exclude SNe 2008ax and 2008bk and focus
our search and analysis on SNe~2007gr and 2011dh.

We find no evidence for GW signals from SNe~2007gr or 2011dh in the
data. Using gravitational waveforms from CCSN
simulations, waveforms generated with phenomenological astrophysical
models, and \textit{ad-hoc} waveforms, we measure the
sensitivity of our search. We show that none of the considered
astrophysical waveforms would likely be detectable at the distances of
SNe~2007gr and 2011dh for the first-generation detector networks. 
Furthermore, even a very strong gravitational wave could potentially be 
missed due to incomplete coverage of the CCSN on-source window by 
the detector network. 
Motivated by this, we provide a statistical approach for model
exclusion by combining observational results for multiple CCSNe. Using
this approach, we quantitatively estimate how increased detector
sensitivity and a larger sample of targeted CCSNe will improve our
ability to rule out the most extreme emission models. This suggests
that observations with second-generation ``Advanced'' interferometers
\cite{aLIGO,aVirgo,kagra} will be able to put interesting constraints
on GW emission of extragalactic CCSN at $D\lesssim 10\,\mathrm{Mpc}$.

The remainder of this paper is structured as follows. In
Section~\ref{sec:ccsne}, we discuss the targeted CCSNe and the
determination of their on-source windows. In Section~\ref{sec:networks},
we describe the detector networks, the coverage of the on-source 
windows with coincident observation, and the data searched. In
Section~\ref{sec:overview}, we present our search methodology and the
waveform models studied. We present the search results in
Section~\ref{sec:results} and conclusions in
Section~\ref{sec:summaries}. 
\section{Targeted Core-Collapse Supernovae}
\label{sec:ccsne}

\begin{table*}[t]
\caption{Core-collapse supernovae selected as triggers for the  gravitational-wave search described in this paper. Distance gives the best current estimate
  for the distance to the host galaxy. $t_1$ and $t_2$ are the UTC
  dates delimiting the on-source window. $\Delta t$ is the temporal
  extent of the on-source window.  iLIGO/Virgo run indicates the data taking campaign during which the supernova explosion was
  observed. Detectors lists the interferometers taking data during at
  least part of the on-source window.  The last column provides the
  relative coverage of the on-source window with science-quality or
  Astrowatch-quality data of at least two detctors.  For SN 2007gr,
  the relative coverage of the on-source window with the most
  sensitive network of four active interferometers is 67\%.  See the
  text in Section~\ref{sec:ccsne} for details and references on the
  supernovae and Section~\ref{sec:networks} for details on the
  detector networks, coverage, and data quality. }
\begin{center}
\begin{tabular}{ll@{\!}rrllrrrr}
\hline
\hline
Identifier & Type &\multicolumn{1}{c}{Host}&\multicolumn{1}{c}{Distance}
& \multicolumn{1}{c}{$t_{1}$} & \multicolumn{1}{c}{$t_{2}$}
& \multicolumn{1}{c}{$\Delta t$} & \multicolumn{1}{c}{iLIGO/Virgo} & \multicolumn{1}{c}{Active}&\multicolumn{1}{c}{Coincident}\\
& &\multicolumn{1}{c}{Galaxy} &\multicolumn{1}{c}{[Mpc]} &\multicolumn{1}{c}{[UTC]} &\multicolumn{1}{c}{[UTC]} & \multicolumn{1}{c}{[days]} &\multicolumn{1}{c}{Run} & \multicolumn{1}{c}{Detectors}&\multicolumn{1}{c}{Coverage}\\
\hline
SN 2007gr & Ic & NGC 1058&$10.55$$\pm$$1.95$ 
& 2007 Aug $10.39$ & 2007 Aug $15.51$ & 5.12 & S5/VSR1 & H1,H2,L1,V1&93\%\\
SN 2008ax & IIb & NGC 4490 &$9.64$$+1.38$$-1.21$ 
& 2008 Mar $2.19$ & 2008 Mar 3.45 & 1.26 & A5 & G1,H2 &8\% \\
SN 2008bk & IIP & NGC 7793 &$3.53$$+0.21$$-0.29$ 
& 2008 Mar $13.50$ & 2008 Mar 25.14 & 11.64 & A5 & G1,H2 &38\%\\
SN 2011dh & IIb & M51 &$8.40$$\pm$$0.70$ 
& 2011 May $30.37$ & 2011 May 31.89 & 1.52 & S6E/VSR4 & G1,V1 &37\%\\
\hline
\hline
\end{tabular} \label{table:triggers}
\end{center}
\end{table*}

For the present search it is important to have an estimate of the time
of core collapse for each supernova.  This time
coincides (within one to a few seconds; e.g.,
\cite{ott:09}) with the time of strongest GW emission. The better the
estimate of the core collapse time, the smaller the \emph{on-source
  window} of detector data that must be searched and the smaller the
confusion background due to non-Gaussian non-stationary detector
noise.

For a Galactic or Magellanic Cloud CCSN, the time of core collapse
would be extremely well determined by the time of arrival of the neutrino
burst that is emitted coincident with the GW signal
\cite{pagliaroli:09}. A very small on-source window of seconds to
minutes could be used for such a special event.

For CCSNe at distances $D\gtrsim 1\,\mathrm{Mpc}$, an observed coincident
neutrino signal is highly unlikely \cite{ikeda:07,leonor:10}. In this
case, the time of core collapse must be inferred based on estimates of
the explosion time, explosion energy, and the radius of the
progenitor. The explosion time is defined as the time at which the
supernova shock breaks out of the stellar surface and the
electromagnetic emission of the supernova begins. Basic information
about the progenitor can be obtained from the lightcurve and spectrum
of the supernova (e.g., \cite{filippenko:97}). Much more information
can be obtained if pre-explosion imaging of the progenitor is
available (e.g., \cite{smartt:09b}). A red supergiant progenitor with
a typical radius of $\sim$$500 - 1500\,R_\odot$ produces a Type IIP
supernova and has an explosion time of $\sim$$1-2\,\mathrm{days}$
after core collapse and a typical explosion energy of
$10^{51}\,\mathrm{erg}$; sub-energetic explosions lead to longer
explosion times (e.g., \cite{kistler:13,matzner:99,morozova:15}).  A
yellow supergiant that has been partially stripped of its
hydrogen-rich envelope, giving rise to a IIb supernova (e.g.,
\cite{bersten:12}), is expected to have a radius of
$\sim$$200-500\,R_\odot$ and an explosion time of $\lesssim
0.5\,\mathrm{days}$ after core collapse
\citep{bersten:12,morozova:15}.  A blue supergiant, giving rise to a
peculiar type IIP supernova (such as SN 1987A), has a radius of
$\lesssim$$100\,R_\odot$ and an explosion time of
$\lesssim$$2-3\,\mathrm{hours}$ after core collapse. A Wolf-Rayet star
progenitor, giving rise to a Type Ib/c supernova, has been stripped of
its hydrogen (and helium) envelope by stellar winds or binary
interactions and has a radius of only a $\mathrm{few}$ to
$\sim$$10\,R_\odot$ and shock breakout occurs within
$\sim$$10-100\,\mathrm{s}$ of core collapse
\cite{kistler:13,matzner:99}.

The breakout of the supernova shock through the surface of the
progenitor star leads to a short-duration high-luminosity burst of
electromagnetic radiation with a spectral peak dependent on the radius
of the progenitor.  The burst from shock-breakout preceeds the rise of
the optical lightcurve which occurs on a timescale of days after shock
breakout (depending, in detail, on the nature of the progenitor star;
\cite{filippenko:97,kasen:09,bersten:12,morozova:15}).

With the exception of very few serendipitous
discoveries of shock breakout bursts (e.g.,
\cite{soderberg:08,gezari:08}), core-collapse supernovae in the
2007--2011 time frame of the present GW search were usually discovered
days after explosion and their explosion time is constrained by one or
multiple of (\emph{i}) the most recent non-detection, i.e., by the
last date of observation of the host galaxy without the supernova present;
(\emph{ii}) by comparison of observed lightcurve and spectra with
those of other supernovae for which the explosion time is well known;
(\emph{iii}) by lightcurve extrapolation \cite{cowen:10}; or,
(\emph{iv}), for type IIP supernovae, via lightcurve modeling using
the expanding photosphere method (EPM; e.g.,
\cite{kirshner:74,dessart:05}).

More than 100 core-collapse supernovae were discovered in the optical
by amateur astronomers and professional astronomers (e.g.,
\cite{snlistweb}) during the S5/S6 iLIGO and the VSR2, VSR3, VSR4 Virgo data
taking periods.  In order to select optically discovered core-collapse
supernovae as triggers for this search, we impose the following
criteria: (\emph{i}) distance from Earth not greater than
$\sim$$10-15\,\mathrm{Mpc}$. Since GWs from core-collapse supernovae
are most likely very weak and because the observable GW amplitude
scales with one-over-distance, nearer events are greatly
favored. (\emph{ii}) A well constrained time of explosion leading to an
uncertainty in the time of core collapse of less than $\sim$2 weeks.
(\emph{iii}) At least partial availability of science-quality data of
coincident observations of more than one interferometer in the
on-source window.

The core-collapse supernovae making these cuts are SN~2007gr,
SN~2008ax, SN~2008bk, and SN~2011dh. Table~\ref{table:triggers}
summarizes key properties of these supernovae and we discuss each in
more detail in the following. 

{\bf SN 2007gr}, a Type Ic supernova, was discovered on 2007 August
15.51 UTC\,\cite{madison:07}. A pre-discovery empty image taken by
KAIT~\cite{filippenko:01} on August 10.44 UTC provides a baseline
constraint on the explosion time. The progenitor of this supernova was
a compact stripped-envelope star
\cite{crockett:08b,mazzali:10,eldridge:13,chen:14} through which the supernova
shock propagated within tens to hundreds of seconds. In order to be
conservative, we add an additional hour to the interval between
discovery and last non-detection and arrive at a GW on-source window
of 2007 August 10.39 UTC to 2007 August 15.51 UTC. The sky location of
SN 2007gr is
$\mathrm{R.A.}\!=02^{\mathrm{h}}43^{\mathrm{m}}27^{\mathrm{s}}.98$, $
\mathrm{Decl.}\!= +37^\circ20{'}44{''}.7$ \cite{madison:07}. The host
galaxy is NGC 1058. Schmidt~\emph{et al.} \cite{schmidt:94} used 
EPM to determine the distance to SN 1969L, which exploded in the same
galaxy. They found $D = (10.6+1.9-1.1)\,\mathrm{Mpc}$. This is broadly
consistent with the more recent Cepheid-based distance estimate of $D
= (9.29\pm0.69)\,\mathrm{Mpc}$ to NGC 925 by
\cite{silbermann:96}. This galaxy is in the same galaxy group as NGC
1058 and thus presumed to be in close proximity. For the purpose of
the present study, we use the conservative combined distance estimate
of $D = (10.55\pm 1.95\,\mathrm{Mpc})$.

\begin{table*}[t]
\caption{Overview of GW interferometer science runs from which we draw
  data for our search. H1 and H2 stand for the LIGO Hanford
  $4$-$\mathrm{km}$ and $2$-$\mathrm{km}$ detectors, respectively. L1
  stands for the LIGO Livingston detector. V1 stands for the Virgo
  detector and G1 stands for the GEO\,600 detector. The duty factor
  column indicates the approximate fraction of science-quality data
  during the observation runs. The coincident duty factor column
  indicates the fraction of time during which at least two detectors
  were taking science-quality data simultaneously. The A5 run was
  classified as \emph{astrowatch} and was not a formal science
  run. The H2 and V1 detectors operated for only part of A5. The Virgo
  VSR1 run was joint with the iLIGO S5 run, the Virgo VSR2 and VSR3 runs
  were joint with the iLIGO S6 run, and the GEO\,600 detector (G1)
  operated in iLIGO run S6E during Virgo run VSR4. When iLIGO and Virgo
  science runs overlap, the coincident duty factor takes into
  account iLIGO, GEO\,600, and Virgo detectors.}  \centering
\begin{tabular}{l@{\hspace{0.3em}}c@{\hspace{1em}}c@{\hspace{1em}}c@{\hspace{1em}}c}
\hline
\hline
Run&Detectors&Run Period&Duty Factors& Coin.\ Duty Factor\\
\hline
S5&H1,H2,L1,G1&2005/11/04--2007/10/01&$\sim$75\% (H1), $\sim$76\% (H2), $\sim$65\% (L1), $\sim$77\% (G1)&$\sim$87\%\\
A5&G1,H2,V1&2007/10/01--2009/05/31& $\sim$81\%(G1), $\sim$18\% (H2), $\sim$5\% (V1)& $\sim$18\%\\
S6&L1,H1,G1&2009/07/07--2010/10/21&$\sim$51\% (H1), $\sim$47\% (L1), $\sim$56\% (G1)&$\sim$67\%\\
S6E&G1&2011/06/03--2011/09/05&$\sim$77\%&$\sim$66\%\\
VSR1/S5&V1&2007/05/18--2007/10/01&$\sim$80\%&$\sim$97\%\\
VSR2/S6&V1&2009/07/07--2010/01/08&$\sim$81\%&$\sim$74\%\\
VSR3/S6&V1&2010/08/11--2010/10/19&$\sim$73\%&$\sim$94\%\\
VSR4/S6E&V1&2011/05/20--2011/09/05&$\sim$78\%&$\sim$62\%\\
\hline
\hline
\end{tabular}
\label{tab:runs}
\end{table*}

{\bf SN 2008ax}, a Type IIb supernova \cite{chornock:08}, was
discovered by KAIT on 2008 March 3.45 UTC\,\cite{mostardi:08}.  The
fortuitous non-detection observation made by Arbour on 2008 March 3.19
UTC \cite{arbour:08}, a mere 6.24 h before the SN discovery, provides
an excellent baseline estimate of the explosion time. Spectral
observations indicate that the progentior of SN 2008ax was almost
completely stripped of its hydrogen envelope, suggesting that is
exploded either as a yellow supergiant or as a Wolf-Rayet
star \cite{chornock:11,crockett:08}. Most recent observations and
phenomenological modeling by \cite{folatelli:15} suggest that the
progenitor was in a binary system and may have had a blue-supergiant
appearance and an extended ($30-40\,R_\odot$) low-density (thus,
low-mass) hydrogen-rich envelope at the time of explosion. To be
conservative, we add an additional day to account for the uncertainty
in shock propagation time and define the GW on-source window as 2008
March 2.19 UTC to 2008 March 3.45 UTC.  The coordinates of SN 2008ax
are $\mathrm{R.A.}\!
=12^{\mathrm{h}}30^{\mathrm{m}}40^{\mathrm{s}}.80$, $
\mathrm{Decl.}\!= +41^\circ38{'}14{''}.5$ \cite{mostardi:08}.  Its
host galaxy is NGC 4490, which together with NGC 4485 forms a pair of
interacting galaxies with a high star formation rate. We adopt the
distance $D = (9.64 +1.38 -1.21)\, \mathrm{Mpc}$ given by
Pastorello~\emph{et al.}~\cite{pastorello:08}

{\bf SN 2008bk}, a Type IIP supernova, was discovered on 2008 March
25.14 UTC \cite{monard:08}.  Its explosion time is poorly constrained
by a pre-explosion image taken on 2008 January 2.74
UTC\,\cite{monard:08}. Morrell \& Stritzinger \cite{morrell:08}
compared a spectrum taken of SN 2008bk on 2008 April 12.4 UTC to a
library of SN spectra \cite{howell:05} and found a best fit to the
spectrum of SN 1999em taken at 36~days after explosion
\cite{morrell:08}. However, the next other spectra available for SN
1999em are from 20 and 75 days after explosion, so the uncertainty of
this result is rather large.  EPM modeling by
Dessart~\cite{dessart:11pc} suggests an explosion time of March
$19.5\pm5$ UTC, which is broadly consistent with the lightcurve data
and hydrodynamical modeling presented in \cite{hamuy:12}. The
progenitor of SN 2008bk was most likely a red supergiant with a radius
of $\sim$$500\,R_\odot$ \cite{maund:14,vandyk:12,mattila:08}, which
suggests an explosion time of $\sim$$1\,\mathrm{day}$ after core
collapse \cite{kistler:13,matzner:99,morozova:15}. Hence, we assume a
conservative on-source window of 2008 March 13.5 UTC to 2008 March
25.14 UTC. The coordinates of SN 2008bk are $\mathrm{R.A.}\!=
23^{\mathrm{h}}57^{\mathrm{m}}50^{\mathrm{s}}.42$, $ \mathrm{Decl.}\!=
-32^\circ 33{'} 21{''}.5$ \cite{li:08_2008bk}.  Its host galaxy is NGC
7793, which is located at a Cepheid-distance $D = (3.44 +0.21
-0.2)\,\mathrm{Mpc}$ \cite{pietrzynski:10}. This distance estimate is
consistent with $D = (3.61 +0.13 -0.14)\,\mathrm{Mpc}$ obtained by
\cite{jacobs:09} based on the tip of the red giant branch method
(e.g., \cite{lee:93}). For the purpose of this study, we use a
conservative averaged estimate of $D = (3.53 +0.21
-0.29)\,\mathrm{Mpc}$.

{\bf SN 2011dh}, a type IIb supernova, has an earliest discovery date  in the literature
of 2011 May 31.893, which was by amateur astronomers
\cite{griga:11,marion:14,ergon:14,horesh:13}. An earlier discovery
date of 2011 May 31.840 is given by Alekseev \cite{rochester2011dh}
and a most recent non-detection by Dwyer on 2011 May 31.365
\cite{rochester2011dh}. The progenitor of SN 2011dh was with high
probability a yellow supergiant star \cite{vandyk:13} with a radius of
a $\mathrm{few}\,100\,R_\odot$
\cite{bersten:12,vinko:12,vandyk:11}. We conservatively estimate an
earliest time of core collapse of a day before the most recent
non-detection by Dwyer and use an on-source window of 2011 May 30.365
to 2011 May 31.893.  SN 2011dh's location is $\mathrm{R.A.}\!=
13^{\mathrm{h}}30^{\mathrm{m}}05^{\mathrm{s}}.12, $ $\mathrm{Decl.}\!=
+47^\circ 10{'} 11{''}.30$ \cite{atel:3406} in the nearby spiral
galaxy M51.  The best estimates for the distance to M51 come
from Vink\'o~\emph{et al.}~\cite{vinko:12}, who give $D =
8.4\pm0.7\,\mathrm{Mpc}$ on the basis of EPM modeling of SN 2005cs and
SN 2011dh. This is in agreement with Feldmeier~\emph{et al.}
\cite{feldmeier:97}, who give $D=8.4\pm0.6\,\mathrm{Mpc}$ on the basis
of planetary nebula luminosity functions. Estimates using surface
brightness variations \cite{tonry:01} or the Tully-Fisher relation
\cite{tully:88} are less reliable, but give a somewhat lower distance
estimates of $D=7.7\pm0.9$ and $D = 7.7\pm1.3$, respectively. We adopt
the conservative distance $D = 8.4\pm0.7\,\mathrm{Mpc}$ for the purpose of
this study.

\section{Detector Networks and Coverage}
\label{sec:networks}

\begin{figure*}[t]
\centering
\includegraphics[width=0.9\textwidth]{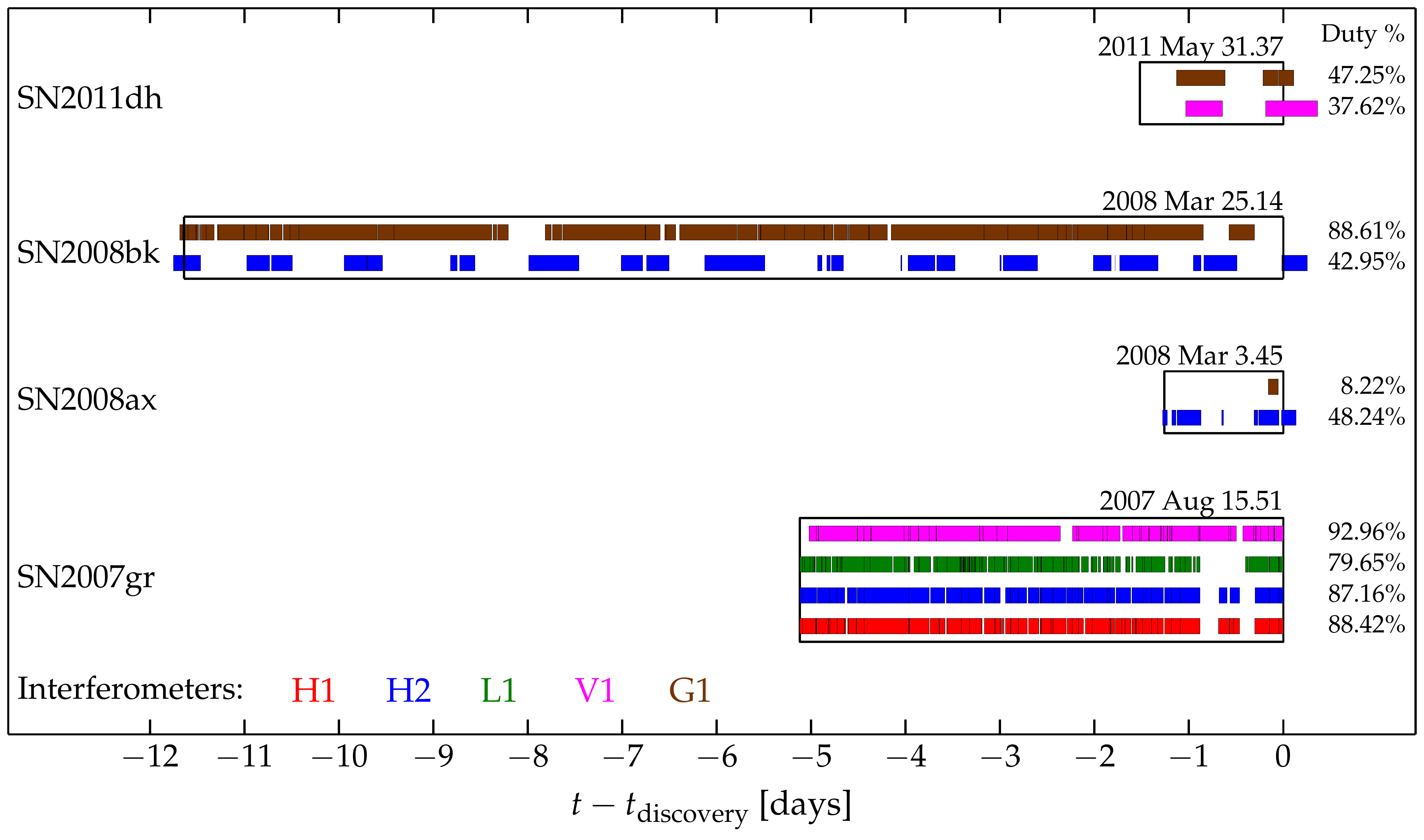}
\caption{On-source windows as defined for the four core-collapse 
  supernovae considered in Section~\ref{sec:ccsne}. The date given
  for each core-collapse supernova is the published date of discovery.
  Overplotted in color are the stretches of time covered with
  science-quality and Astrowatch-quality data of the various GW
  interferometers.  The percentages given for each core-collapse
  supernova and interferometer is the fractional coverage of the
  on-source window with science or astrowatch data by that
  interferometer. See
  Table~\ref{table:triggers} and Sections~\ref{sec:ccsne} and
  \ref{sec:networks} for details.}
\label{fig:coverage}
\end{figure*}

This search employs data from the $4\,\mathrm{km}$ LIGO Hanford, WA
and LIGO Livingston, LA interferometers (denoted \textbf{H1} and
\textbf{L1}, respectively), from the $2\,\mathrm{km}$ LIGO Hanford, WA
interferometer (denoted as \textbf{H2}), from the $0.6\,\mathrm{km}$
GEO 600 detector near Hannover, Germany (denoted as \textbf{G1}), and
from the $3\,\mathrm{km}$ Virgo interferometer near Cascina, Italy
(denoted as \textbf{V1}).

Table~\ref{tab:runs} lists the various GW interferometer data taking
periods (``runs'') in the 2005--2011 time frame from which we draw
data for our search. The table also provides the duty factor and
\emph{coincident} duty factor of the GW interferometers. The duty
factor is the fraction of the run time a given detector was
taking science-quality data. The coincident duty factor is the
fraction of the run time at least two detectors were taking science
quality data. The coincident duty factor is most relevant for GW
searches like ours that require data from at least two detectors to
reject candidate events that are due to non-Gaussian instrumental or
environmental noise artifacts (``glitches'') but can mimic real
signals in shape and time-frequency content (see, e.g.,
\cite{LIGO,ligo_burst_s5y1:09}).

One notes from Table~\ref{tab:runs} that the duty factor for the
first-generation interferometers was typically $\lesssim50-80\%$.
The relatively low duty factors are due to a combination of 
environmental causes (such as distant
earthquakes causing loss of interferometer lock) and interruptions for
detector commissioning or maintenance.

\begin{figure}[t]
\centering
\includegraphics[width=1.0\columnwidth]{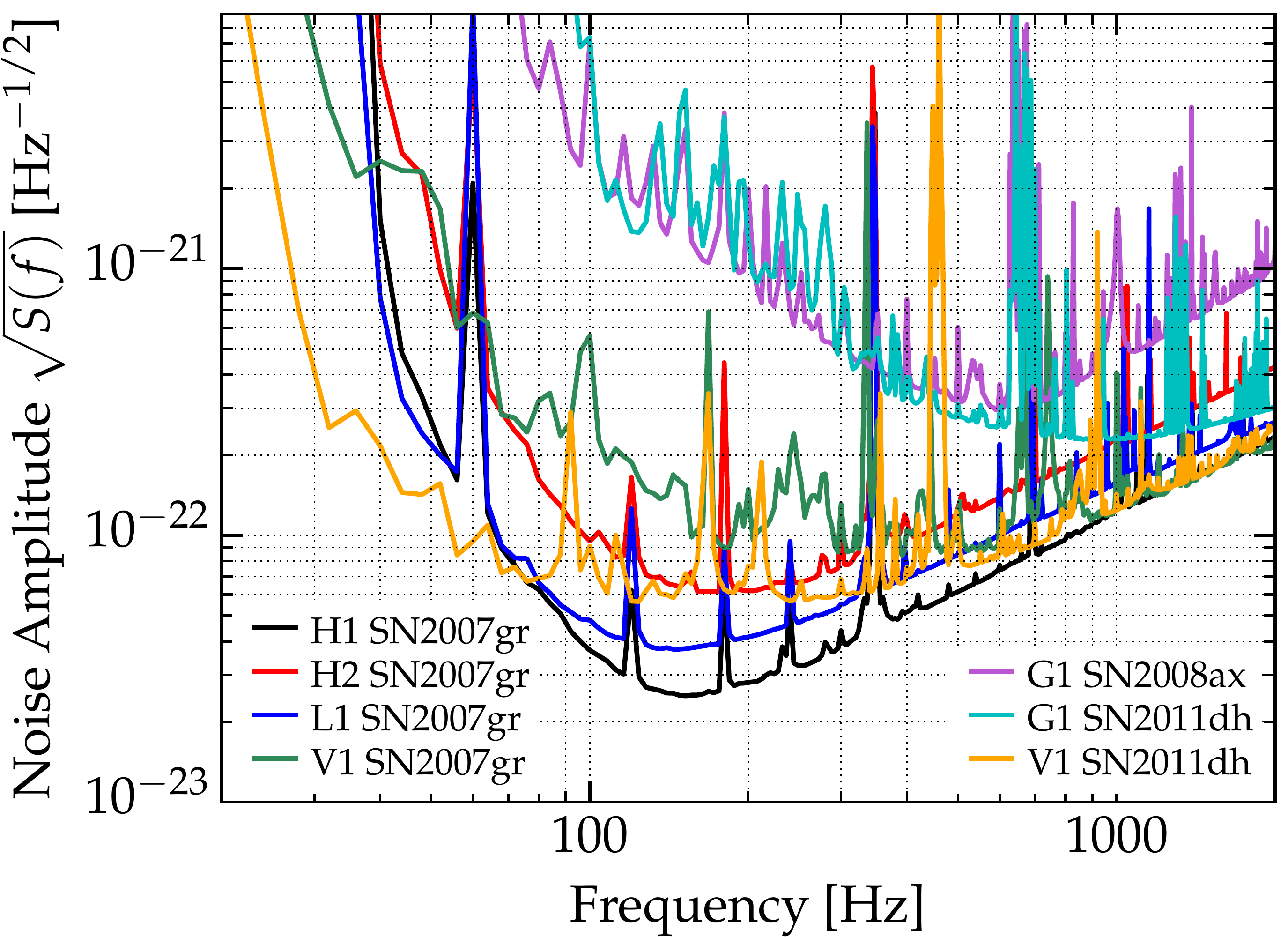}
\caption{Noise amplitude spectral densities of the GW interferometers
  whose data are analyzed for SNe 2007gr and 2011dh (see
  Section~\ref{sec:networks}). The curves are the results of averaging
  $1/S(f)$ over the on-source windows of the SNe (see
  Table~\ref{table:triggers}). We plot the G1 noise spectrum also for
  SN~2008ax to demonstrate the improvement in high-frequency
  sensitivity due to GEO-HF \cite{grote:13squeeze} for SN~2011dh.}
\label{fig:noise}
\end{figure}

The CCSNe targeted by this search and described in 
Section~\ref{sec:ccsne} are the only 2007--2011 CCSNe located within
$D\lesssim 10-15 \,\mathrm{Mpc}$ for which well-defined on-source
windows exist and which  are also covered by extended stretches of coincident
observations of at least two interferometers. In
Figure~\ref{fig:coverage}, we depict the on-source windows for 
SNe 2007gr, 2008ax, 2008bk, and 2011dh. We indicate with regions of
different color times during which the various interferometers were
collecting data. 

SN 2007gr exploded during the S5/VSR1 joint run between the iLIGO, GEO
600, and Virgo detectors. It has the best coverage of all considered
CCSNe: 93\% of its on-source window are covered by science-quality
data from at least two of H1, H2, L1, and V1.  We search for GWs from
SN 2007gr at times when data from the following detector networks are
available: H1H2L1V1, H1H2L1, H1H2V1, H1H2, L1V1.  The G1 detector was
also taking data during SN 2007gr's on-source window, but since its
sensitivity was much lower than that of the other detectors, we do not
analyze G1 data for SN 2007gr.

SNe 2008ax and 2008bk exploded in the A5 \emph{astrowatch} run between
the S5 and S6 iLIGO science runs (cf.~Table~\ref{tab:runs}).  Only
the G1 and H2 detectors were operating at sensitivities much lower
than those of the $4$-km L1 and H1 and the $3$-km V1 detectors.  The
coincident duty factor for SN~2008ax is only 8\% while that for
SN~2008bk is 38\%. Preliminary analysis of the available coincident GW
data showed that due to a combination of low duty factors and low
detector sensitivity, the overall sensitivity to GWs from these CCSNe
was much lower than for SNe~2007gr and 2011dh. Because of this, we
exclude SNe~2008ax and 2008bk from the analysis presented in the rest
of this paper.

SN 2011dh exploded a few days before the start of the S6E/VSR4 run
during which the V1 and G1 interferometers were operating
(cf.~Table~\ref{tab:runs}). G1 was operating in GEO-HF mode
\cite{grote:13squeeze} that improved its high-frequency ($f \gtrsim
1\,\mathrm{kHz}$) sensitivity to within a factor of two of V1's
sensitivity.  While not officially in a science run during the
SN~2011dh on-source window, both G1 and V1 were operating and
collecting data that passed the data quality standards necessary for
being classified as science-quality data (e.g.,
\cite{VSRdetchar:12,S6detchar:15,mciver:12}). The coincident G1V1 duty
factor is 37\% for SN 2011dh.

In Figure~\ref{fig:noise}, we plot the one-side noise amplitude spectral 
densities of each detector averaged over the on-source windows of
SNe~2007gr and 2011dh. In order to demonstrate the high-frequency
improvement in the 2011 G1 detector, we also plot the G1 noise 
spectral density for SN~2008ax for comparison.

\section{Search Methodology}\label{sec:overview}

\begin{figure*}[t]
  \begin{minipage}[l][][t]{0.495\textwidth}
    \vspace*{\fill}
	\flushleft
    \includegraphics[width=0.97\linewidth]{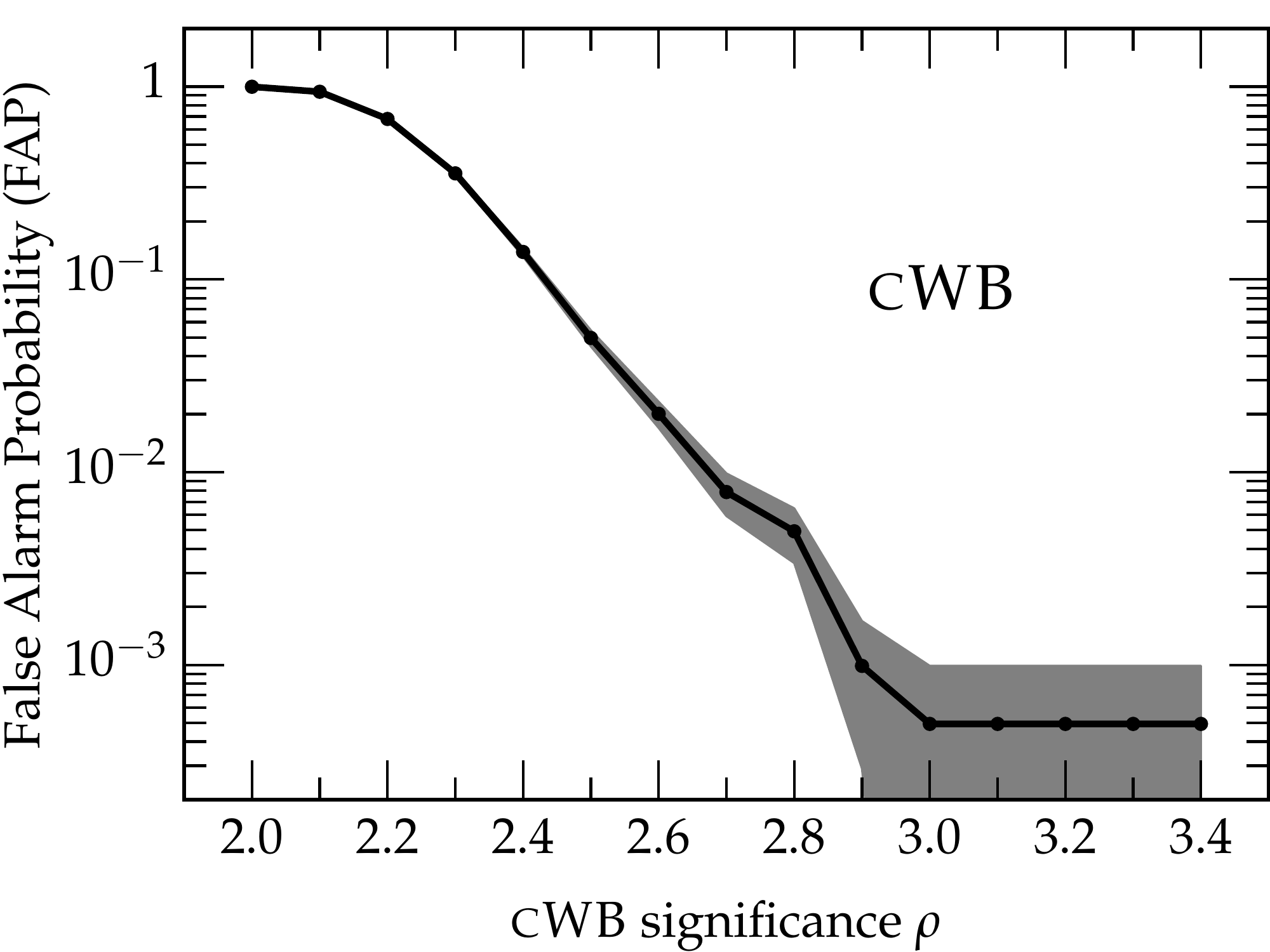}
  \end{minipage}
  \begin{minipage}[r][][t]{0.495\textwidth}
    \vspace*{\fill}
	\flushright
    \includegraphics[width=0.97\linewidth]{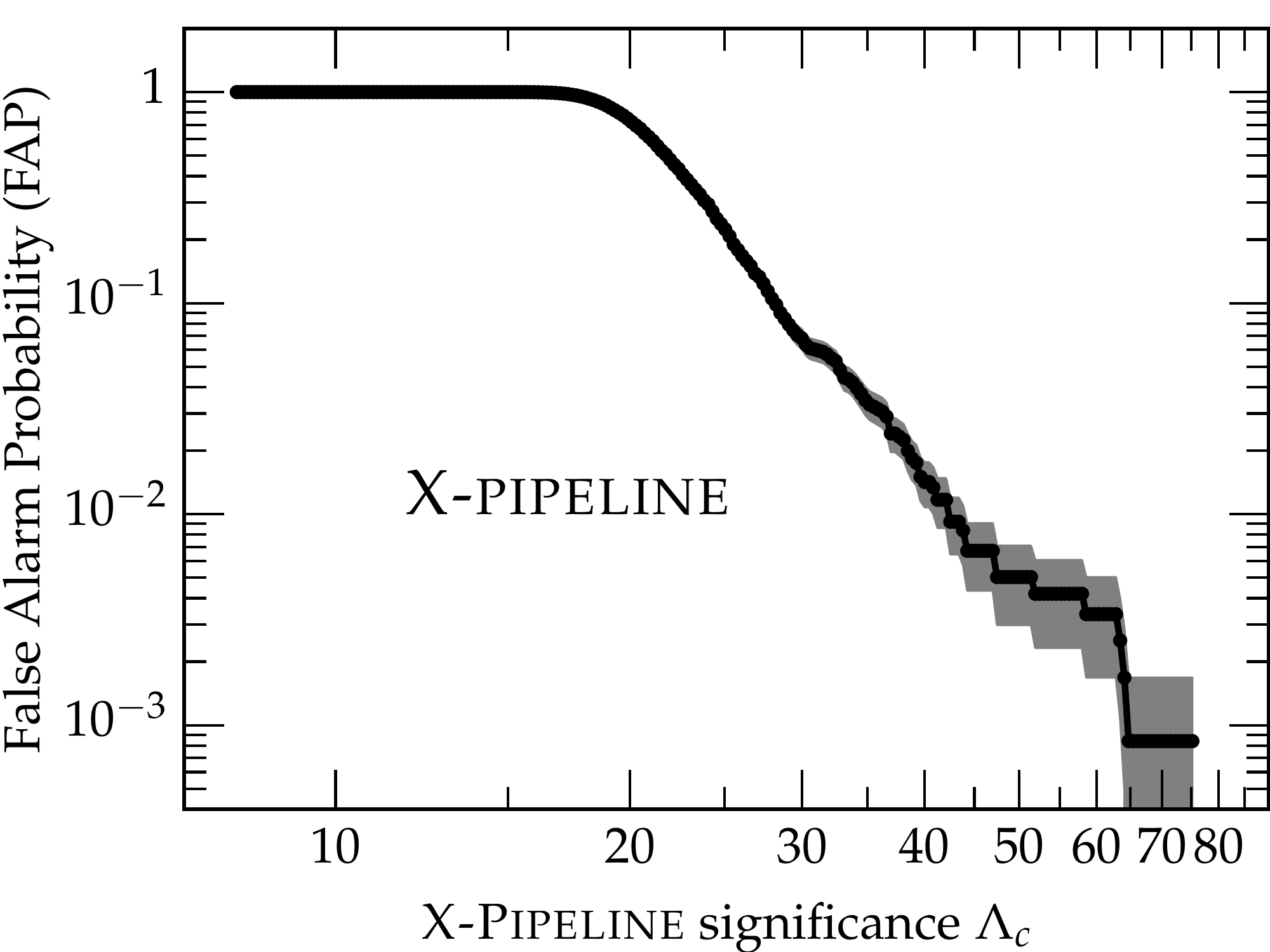}
  \end{minipage} 
  \caption{False Alarm Probability [FAP,
      Equation~(\ref{eqn:fap})] distributions of the background events
      for SN~2007gr and the H1H2L1V1 detector network
      (cf.\ Section~\ref{sec:networks}).  The FAP indicates the
      probability that an event of a given ``loudness''
      (significance) is consistent with background noise.  The
      left panel shows the FAP distribution determined by the
      {\cwb} pipeline as a function of its loudness measure, $\rho$,
      (see \cite{klimenko:08} for details).  The right panel depicts
      the same for {\xp} as a function of its loudness measure,
      $\Lambda_c$, (see \cite{sutton:10,Was:2012zq} for
      details). The shaded regions indicate $1-\sigma$ error
      estimates for the FAP. }
  \label{fig:bckgrddist}
\end{figure*}

Two search algorithms are employed in this study: {\xp}
\cite{sutton:10,Was:2012zq} and Coherent WaveBurst ({\cwb})
\cite{klimenko:08}.  Neither algorithm requires detailed assumptions
about the GW morphology and both look for subsecond GW transients in
the frequency band 60\,Hz to 2000\,Hz. This is the most sensitive band
of the detector network, where the amplitude of the noise spectrum of
the most sensitive detector is within about an order of magnitude of
its minimum.  This band also encompasses most models for GW emission from
CCSNe (cf.~\cite{ott:09,kotake:13review,fryer:11}).  The benefit of
having two independent algorithms is that they can act as a cross
check for outstanding events.  Furthermore, sensitivity studies using
simulated GWs show some complementarity in the signals detected by
each pipeline; this is discussed further in Section~\ref{sec:results}.

The two algorithms process the data independently to identify potential GW 
events for each supernova and network combination. Each algorithm assigns a 
``loudness'' measure to each event; these are described in more detail below. 
The two algorithms also evaluate measures of signal consistency across 
different interferometers and apply thresholds on these measures (called 
coherence tests) to reject background noise events.
We also reject events that occur at times of environmental noise 
disturbances that are known to be correlated with transients in the GW data 
via well-established physical mechanisms; these so-called ``category 2'' data 
quality cuts are described in~\cite{S5y2Burst}.

The most important measure of an event's significance is its false
alarm rate (FAR): the rate at which the background noise produces
events of equal or higher loudness than events that pass all coherent
tests and data quality cuts.  Each pipeline estimates the FAR using
background events generated by repeating the analysis on time-shifted
data --- the data from the different detectors are offset in time, in
typical increments of $\sim 1$\,s. The shifts remove the chance of
drawing a sub-second GW transient into the background sample since the
largest time of flight between the LIGO and Virgo sites is 27
milliseconds (between H1 and V1).  To accumulate a sufficient sampling
of rare background events, this shifting procedure is performed
thousands of times without repeating the same relative time shifts
among detectors.
Given a total duration $T_\mathrm{off}$ of off-source (time-shifted) data, 
the smallest false alarm rate that can be measured is $1/T_\mathrm{off}$. 

On-source events from each combination of CCSN, detector 
network, and pipeline are assigned a FAR using the time-slide 
background from that combination only. 
The event lists from the different CCSNe, detector networks, and pipelines 
are then combined and the events ranked by their FAR. 
The event with lowest FAR is termed the {\it loudest event}.

In order for the loudest event to be considered as a GW detection 
it must have a False Alarm Probability (FAP) low enough that it is implausible to have been caused 
by background noise. Given a FAR value $R$, 
the probability $p(R)$ of noise producing one or more events of FAR less than 
or equal to $R$ during one or more CCSN on-source windows of total duration $T_\mathrm{on}$ is 
\begin{equation}
  \label{eqn:fap}
p = 1-\exp{(-R T_\mathrm{on})} \, .
\end{equation}
The smallest such false alarm probability (FAP) that can be measured 
given an off-source (time-shifted) data duration $T_\mathrm{off}$ is 
approximately $T_\mathrm{on}/T_\mathrm{off}$. Several thousand time 
shifts are therefore sufficient to measure FAP values of $O(10^{-3})$.  
We require a FAP below 0.001, which exceeds 3-$\sigma$ confidence, 
in order to consider an event to be a possible GW detection candidate.
Figure~\ref{fig:bckgrddist} shows examples of the FAP as a function 
of event loudness for {\cwb} and {\xp} for the H1H2L1V1 network 
during the SN~2007gr on-source window.

If no on-source events have a FAP low enough to be considered 
GW candidates, then we can set upper limits on the strength of any 
GW emission by the CCSNe. 
This is done by adding to the data simulated GW signals of various 
amplitudes (or equivalently sources at various distances) and repeating 
the analysis. For each amplitude or distance we measure the fraction 
of simulations that produce an event in at least one pipeline with FAP 
lower than the loudest on-source event, and which survive our coherence 
tests and data quality cuts; this fraction is the \textit{detection efficiency} 
of the search.

\subsection{Coherent WaveBurst}
\label{sec:cwb}

The {\cwb} \cite{klimenko:08} analysis is performed as described in
\cite{ligo_burst_s5y1:09}, and it is based on computing a constrained
likelihood function. In brief: each detector data stream is decomposed
into 6 different wavelet decompositions (each one with different time and frequency resolutions).
The data are whitened, and the largest 0.1 percent of wavelet
magnitudes in each frequency bin and decomposition for each
interferometer are retained (we call these ``black pixels''). We also retain ``halo'' pixels, which are those that surround each
black pixel. 
In order to choose pixels that are more likely related to a 
GW transient ({\it candidate event}) we identify clusters of them. Once all of the wavelet decompositions 
are projected into the same time frequency plane, clusters are defined as sets of contiguous retained pixels (black or halo).
Only the pixels involved in a cluster are used in the subsequent calculation of the 
likelihood. These clusters also need to be consistent between
interferometers for the tested direction of arrival.  For each cluster
of wavelets, a Gaussian likelihood function is computed, where the
unknown GW is reconstructed with a maximum-likelihood
estimator. 

The likelihood analysis is repeated over a grid of sky
positions covering the range of possible directions to the GW source.
Since the sky location of each of the analyzed CCSNe is well known, we
could choose to apply this procedure only for the known CCSN sky
location.  However, the detector noise occasionally forces the {\cwb}
likelihood to peak in a sky location away from the true sky
location. As a consequence, some real GW events could be assigned a smaller
likelihood value, lowering the capability to detect them.
Because of this, we consider triggers that fall within an error region
of $0.4$~degrees of the known CCSN sky location and that pass the
significance threshold, even if they are not at the peak of the {\cwb}
reconstructed sky position likelihood. The $0.4$~degree region is
determined empirically by trade-off studies between detection
efficiency and FAR.

For SN~2011dh, the noise spectra were very different for the G1 and V1
detectors, with the consequence that the network effectively had only
one detector at frequencies up to several hundred Hz, and therefore
location reconstruction was very poor. As a consequence we decided to
scan the entire sky for candidate events for this CCSN.

The events reported for a given network configuration are internally
ranked for detection purposes by {\cwb} using the coherent network amplitude statistic $\rho$
defined in \cite{drago:11}. 
Other constraints related to the degree of similarity of the reconstructed 
signal across different interferometers (the ``network correlation coefficient'' $cc$) 
and the ability of the network to reconstruct both polarizations of the GW signal 
(called {\it{regulators}}) are applied to reject background events; 
these are also described in \cite{drago:11}.

\subsection{{\xp}}

In the {\xp} \cite{sutton:10,Was:2012zq,chatterji:06} analysis, the
detector data are first whitened, then Fourier transformed.  A total
energy map is made by summing the spectrogram for each detector, and
``hot'' pixels are identified as the 1\% in each detector with the
largest total energy.  Hot pixels that share an edge or vertex
(nearest neighbors and next-nearest neighbors) are clustered. For each
cluster, the raw time-frequency maps are recombined in a number of
linear combinations designed to give maximum-likelihood estimates of
various GW polarizations given the known sky position of the CCSN. The
energy in each combination is recorded for each cluster, along with
various time-frequency properties of the cluster.  The procedure is
repeated using a series of Fourier transform lengths from $1/4\,$s,
$1/8\,$s, \ldots $1/128\,$s.  Clusters are ranked internally using a
Bayesian-inspired estimate $\Lambda_c$ of the likelihood
ratio for a circularly polarized GW, marginalized over the unknown GW
amplitude $\sigma_h$ with a Jeffreys (logarithmic) prior
$\sigma_h^{-1}$; see \cite{Searle:2007uv,searle:09,Was:2012zq} for
details.

When clusters from different Fourier transform lengths overlap in
time-frequency, the cluster with the largest likelihood $\Lambda_c$ is
retained and the rest are discarded.  Finally, a post-processing
algorithm tunes and applies a series of pass/fail tests to reject
events due to background noise; these tests are based on measures of
correlation between the detectors for each cluster.  The tuning of
these tests is described in detail in \cite{sutton:10}.  For more
details see also \cite{gossan:16}.

\begin{table*}
\caption{
  Injection waveforms from detailed multi-dimensional CCSN
  simulations described in the text.  
  For each waveform, we give the
  emission type, journal reference, waveform identifier,
  angle-averaged root-sum-squared strain $h_\mathrm{rss}$, 
  the frequency $f_\mathrm{peak}$ at which the GW energy spectrum peaks, the
  emitted GW energy $E_\mathrm{GW}$, and available polarizations.
  See \cite{gossan:16,szczepanczykdcc:15} for details.
  }
\begin{tabular}{lclcrcc}
\hline
\hline
\multicolumn{1}{c}{Emission Type}
&\multicolumn{1}{c}{Ref.}
&\multicolumn{1}{c}{Waveform Identifier}
&\multicolumn{1}{c}{$h_\mathrm{rss}$}
&\multicolumn{1}{c}{$f_\mathrm{peak}$}
&\multicolumn{1}{c}{$E_\mathrm{GW}$}
&\multicolumn{1}{c}{Polarizations} \\
&&
&\multicolumn{1}{c}{$[10^{-22} @ 10\,\mathrm{kpc}]$}
&\multicolumn{1}{c}{$[\mathrm{Hz}]$}
&\multicolumn{1}{c}{$[10^{-9} M_{\odot} c^{2}]$}\\
\hline
Rotating Core Collapse
&\cite{dimmelmeier:08}
&Dim1-s15A2O05ls
&1.052
&774
&\phantom{0}\phantom{0}7.685
&$+$\\
Rotating Core Collapse
&\cite{dimmelmeier:08}
&Dim2-s15A2O09ls
&1.803
&753
&\phantom{0}27.873
&$+$\\
Rotating Core Collapse
&\cite{dimmelmeier:08}
&Dim3-s15A3O15ls
&2.690
&237
&\phantom{0}\phantom{0}1.380
&$+$\\
\hline
2D Convection
&\cite{yakunin:10}
&Yakunin-s15
&1.889
&888
&\phantom{0}\phantom{0}9.079
&$+$\\
\hline
3D Convection
&\cite{mueller:e12}
&M\"uller1-L15-3
&1.655
&150
&\phantom{0}$3.741\times10^{-2}$
&$+$, $\times$\\
3D Convection
&\cite{mueller:e12}
&M\"uller2-N20-2
&3.852
&176
&\phantom{0}$4.370\times10^{-2}$
&$+$, $\times$\\
3D Convection
&\cite{mueller:e12}
&M\"uller3-W15-4
&1.093
&204
&\phantom{0}$3.247\times10^{-2}$
&$+$, $\times$\\
\hline
Protoneutron Star Pulsations
&\cite{ott:09}
&Ott-s15
&5.465
&971
&429.946
&$+$\\
\hline
\hline
\end{tabular}
\label{tab:numwaveforms}
\end{table*}

\subsection{Simulated Signals and Search Sensitivity}
\label{sec:simulations}

An important aspect of the GW search presented in this study is to
understand how sensitive the GW detector networks are to GWs emitted
by the considered CCSNe. We establish sensitivity via Monte Carlo
simulation in the following way:

\begin{enumerate}
\item We determine the loudest event in the on-source window that is
  consistent with the CCSN location (and the angular uncertainty
of the search algorithms).
\item We ``inject'' (add) theoretical waveforms scaled to a specific distance
  (or emitted GW energy) every $100\,\mathrm{s}$ plus a randomly
  selected time in $[-10,10]\,\mathrm{s}$ into the time-shifted
  background data. We compare the loudness of the recovered injections
  with the loudest on-source event and record the fraction of the
  injections that passed the coherent tests and data quality cuts and 
  were louder than the loudest on-source event. This
  fraction is the \emph{detection efficiency}.
\item We repeat step (2) for a range of distances (or emitted GW energies)
  to determine the detection efficiency as a function of distance (or
  emitted GW energy).
\end{enumerate}
We refer the reader to~\cite{gossan:16} for more 
details on the injection procedure. 

In this paper, we employ three classes of GW signals for our Monte
Carlo studies: (1) representative waveforms from detailed
multi-dimensional (2D axisymmetric or 3D) CCSN simulations; (2)
semi-analytic phenomenological waveforms of plausible but extreme
emission scenarios; and (3) \textit{ad-hoc} waveform models with tuneable frequency
content and amplitude to establish upper limits on the energy emitted
in GWs at a fixed CCSN distance.
We briefly summarize the nature of these waveforms below.
We list all employed waveforms in
Tables~\ref{tab:numwaveforms} and \ref{tab:phenowaveforms} and
summarize their key emission metrics. In particular, we provide the
angle-averaged root-sum-squared GW strain,
\begin{equation}
\label{eq:hrss}
\hrss=  \sqrt{\int \left\langle h^2_+(t) + h^2_\times(t) \right\rangle_\Omega \mathrm{d}t}\,\, ,
\end{equation}
and the energy $E_\mathrm{GW}$ emitted in GWs, using the expressions given
in \cite{gossan:16}.

\subsubsection{Waveforms from Multi-Dimensional CCSN Simulations}

Rotation leads to a natural axisymmetric quadrupole (oblate)
deformation of the collapsing core. The tremendous acceleration at
core bounce and proto-neutron star formation results in a strong
linearly-polarized burst of GWs followed by a ring-down
signal. Rotating core collapse is the most extensively studied GW
emission process in the CCSN context (see, e.g.,
\cite{mueller:82,zwerger:97,dimmelmeier:02,kotake:03,ott:04,dimmelmeier:08,ott:07prl,abdikamalov:14}
and \cite{ott:09,kotake:13review,fryer:11} for reviews).  For the
purpose of this study, we select three representative rotating core
collapse waveforms from the 2D general-relativistic study of
Dimmelmeier~\emph{et al.}~\cite{dimmelmeier:08}. The simulations
producing these waveforms used the core of a $15$-$M_\odot$ progenitor
star and the Lattimer-Swesty nuclear equation of state
\cite{lseos:91}. The waveforms are enumerated by Dim1--Dim3 prefixes
and are listed in Table~\ref{tab:numwaveforms}. They span the range
from moderate rotation (Dim1-s15A2O05ls) to extremely rapid rotation
(Dim3-s15A3O15ls).  See \cite{dimmelmeier:08} for details on the
collapse dynamics and GW emission.

\begin{table*}[t] 
\caption{
  Injection waveforms from phenomenological and \textit{ad-hoc} emission
  models described in the text. 
  For each waveform, we give the
  emission type, journal reference, waveform identifier,
  angle-averaged root-sum-squared strain $h_\mathrm{rss}$, 
  the frequency $f_\mathrm{peak}$ at which the GW energy spectrum peaks, the
  emitted GW energy $E_\mathrm{GW}$, and available polarizations.
  See \cite{gossan:16,szczepanczykdcc:15} for details.
  As sine-Gaussian waveforms are \textit{ad-hoc}, they can be 
  rescaled arbitrarily and do not have a defined 
  physical distance or $E_\mathrm{GW}$ value.
  } 
\begin{tabular}{lclcrcc}
\hline
\hline
\multicolumn{1}{c}{Emission Type}
&\multicolumn{1}{c}{Ref.}
&\multicolumn{1}{c}{Waveform Identifier}
&\multicolumn{1}{c}{$\hrss$}
&\multicolumn{1}{c}{$f_\mathrm{peak}$}
&\multicolumn{1}{c}{$E_\mathrm{GW}$}
&\multicolumn{1}{c}{Polarizations}\\
&&
&\multicolumn{1}{c}{$[10^{-20} @ 10\,\mathrm{kpc}]$}
&\multicolumn{1}{c}{$[\mathrm{Hz}]$}
&\multicolumn{1}{c}{$[M_{\odot} c^{2}]$}\\
\hline
Long-lasting Bar Mode
&\cite{ott:10dcc}
&LB1-M0.2L60R10f400t100
&\phantom{0}1.480
&800
&\phantom{0}$2.984\times10^{-4}$
&$+,\times$\\
Long-lasting Bar Mode
&\cite{ott:10dcc}
&LB2-M0.2L60R10f400t1000
&\phantom{0}4.682
&800
&\phantom{0}$2.979\times10^{-3}$
&$+,\times$\\
Long-lasting Bar Mode
&\cite{ott:10dcc}
&LB3-M0.2L60R10f800t100
&\phantom{0}5.920
&1600
&\phantom{0}$1.902\times10^{-2}$
&$+,\times$\\
Long-lasting Bar Mode
&\cite{ott:10dcc}
&LB4-M1.0L60R10f400t100
&\phantom{0}7.398
&800
&\phantom{0}$7.459\times10^{-3}$
&$+,\times$\\
Long-lasting Bar Mode
&\cite{ott:10dcc}
&LB5-M1.0L60R10f400t1000
&23.411
&800
&\phantom{0}$7.448\times10^{-2}$
&$+,\times$\\
Long-lasting Bar Mode
&\cite{ott:10dcc}
&LB6-M1.0L60R10f800t25
&14.777
&1601
&\phantom{0}$1.184\times10^{-1}$
&$+,\times$\\
\hline
Torus Fragmentation Instability
& \cite{piro:07}
&Piro1-M5.0$\eta$0.3
&\phantom{0}2.550
&2035
&\phantom{0}$6.773\times10^{-4}$
&$+,\times$\\
Torus Fragmentation Instability
& \cite{piro:07}
&Piro2-M5.0$\eta$0.6
&\phantom{0}9.936
&1987
&\phantom{0}$1.027\times10^{-2}$
&$+,\times$\\
Torus Fragmentation Instability
& \cite{piro:07}
&Piro3-M10.0$\eta$0.3
&\phantom{0}7.208
&2033
&\phantom{0}$4.988\times10^{-3}$
&$+,\times$\\
Torus Fragmentation Instability
& \cite{piro:07}
&Piro4-M10.0$\eta$0.6
&28.084
&2041
&\phantom{0}$7.450\times10^{-2}$
&$+,\times$\\

\hline
sine-Gaussian
&\cite{abadie:12s6burst}
&SG1-235HzQ8d9linear
&---
&235
&---
&$+$\\
sine-Gaussian
&\cite{abadie:12s6burst}
&SG2-1304HzQ8d9linear
&---
&1304
&---
&$+$\\
sine-Gaussian
&\cite{abadie:12s6burst}
&SG3-235HzQ8d9elliptical
&---
&235
&---
&$+,\times$\\
sine-Gaussian
&\cite{abadie:12s6burst}
&SG4-1304HzQ8d9elliptical
&---
&1304
&---
&$+,\times$\\
\hline
\hline
\end{tabular}
\label{tab:phenowaveforms}
\end{table*}

In non-rotating or slowly rotating CCSNe, neutrino-driven convection
and the standing accretion shock instability (SASI) are expected to
dominate the GW emission. GWs from convection/SASI have also been
extensively studied in 2D (e.g.,
\cite{jm:97,mueller:04,kotake:07a,murphy:09,kotake:09,marek:09b,mueller:13gw,yakunin:10,yakunin:15})
and more recently also in 3D \cite{mueller:e12,ott:13a}. For the
present study, we select a waveform from a 2D Newtonian (+
relativistic corrections) radiation-hydrodynamics simulation of a CCSN
in a $15$-$M_\odot$ progenitor by Yakunin~\emph{et
  al.}~\cite{yakunin:10}. This waveform and its key emission metrics
are listed as Yakunin-s15 in Table~\ref{tab:numwaveforms}. Note that
since the simulation producing this waveform was axisymmetric, only
the $+$ polarization is available.

CCSNe in Nature are 3D and produce both GW polarizations ($h_+$ and
$h_\times$). Only a few GW signals from 3D simulations are presently
available.  We draw three waveforms from the work of M\"uller~\emph{et
  al}.~\cite{mueller:e12}. These and their key GW emission
characteristics are listed with M\"uller1--M\"uller3 prefixes in
Table~\ref{tab:numwaveforms}. Waveforms M\"uller1-L15-3 and
M\"uller2-W15-4 are from simulations using two different progenitor
models for a $15$-$M_\odot$ star. Waveform M\"uller2-N20-2 is from a
simulation of a CCSN in a $20$-$M_\odot$ star. Note that the
simulations of M\"uller~\emph{et al.}~\cite{mueller:e12} employed an
\textit{ad-hoc} inner boundary at multiple tens of kilometers. This prevented
decelerating convective plumes from reaching small radii and high
velocities. As a consequence, the overall GW emission in these
simulations peaks at lower frequencies than in simulations that do not
employ an inner boundary 
(cf.~\cite{yakunin:10,yakunin:15,mueller:13gw,ott:13a}).

In some 2D CCSN simulations \cite{burrows:06,burrows:07a}, strong
excitations of an $\ell = 1$ \emph{g}-mode (an oscillation mode with
gravity as its restoring force) were observed. These oscillations were
found to be highly non-linear and to couple to GW-emitting $\ell = 2$
modes. The result is a strong burst of GWs that lasts for the duration
of the large-amplitude mode excitation, possibly for hundreds of
milliseconds \cite{ott:09,ott:06prl}. More recent simulations do not
find such strong \emph{g}-mode excitations (e.g.,
\cite{marek:09,mueller:13gw}). We nevertheless include here one
waveform from the simulations of \cite{burrows:07a} that was
reported by Ott~\cite{ott:09}. This waveform is from a simulation with
a $15$-$M_\odot$ progenitor and is denoted as Ott-s15 in
Table~\ref{tab:numwaveforms}.

\subsubsection{Phenomenological Waveform Models}

In the context of rapidly rotating core collapse, various
non-axisymmetric instabilities can deform the proto-neutron star into
a tri-axial (``bar'') shape (e.g.,
\cite{lai:95,brown:01,shibata:05,rmr:98,rotinst:05,ott:07prl,scheidegger:10b}),
potentially leading to extended ($\sim$$10\,\mathrm{ms} -
\,\mathrm{few}\,\mathrm{s}$) and energetic GW emission. This emission
occurs at twice the proto-neutron star spin frequency and with
amplitude dependent on the magnitude of the bar deformation
\cite{fryer:02,scheidegger:10b,ott:07prl}. Since few long-term 3D
simulations are available, we resort to the simple phenomenological
bar model described in \cite{ott:10dcc}. Its parameters are the length
of the bar deformation, $L$, in km, its radius, $R$, in km, the mass,
$M$, in $M_\odot$, involved in the deformation, the spin frequency,
$f$, and the duration, $t$, of the deformation. We select six
waveforms as representative examples. We sample the potential
parameter space by chosing $M = \{0.2,1.0\}\,M_\odot$, $f =
\{400,800\}\,\mathrm{Hz}$, and $t = \{25, 100,
1000\}\,\mathrm{ms}$. We list these waveforms as ``Long-lasting Bar
Mode'' in Table~\ref{tab:phenowaveforms} and enumerate them as
LB1--LB6. The employed model parameters are encoded in the full
waveform name.  One notes from Table~\ref{tab:phenowaveforms} that the
strength of the bar-mode GW emission is orders of magnitude greater
than that of any of the waveforms computed from detailed
multi-dimensional simulations listed in Table~\ref{tab:numwaveforms}.
We emphasize that the phenomenological bar-mode waveforms should be
considered as being at the extreme end of plausible GW emission
scenarios. Theoretical considerations (e.g., \cite{ott:09}) suggest
that such strong emission is unlikely to obtain in CCSNe.
Observationally, however, having this emission in one 
or all of the CCSNe has not been ruled out.

We also consider the phenomenological waveform model proposed by Piro
\& Pfahl \cite{piro:07}. They considered the formation of a dense
self-gravitating $M_\odot$-scale fragment in a thick accretion torus
around a black hole in the context of collapsar-type gamma-ray
bursts. The fragment is driven toward the black hole by a combination
of viscous torques and energetic GW emission. This is an extreme but
plausible scenario. We generate injection waveforms from this model
using the implementation described in \cite{santamaria:11dcc}. The
model has the following parameters: mass $M_\mathrm{BH}$ of the black
hole in $M_\odot$, a spatially constant geometrical parameter
controlling the torus thickness, $\eta = H/r$, where $H$ is the disk
scale height and $r$ is the local radius, a scale factor for the
fragment mass (fixed at $0.2$), the value of the phenomenological
$\alpha$-viscosity (fixed at $\alpha = 0.1$), and a starting radius
that we fix to be $100 r_g = 100 GM_\mathrm{BH}/c^2$. We employ four
waveforms, probing black hole masses $M_\mathrm{BH} =
\{5,10\}\,M_\odot$ and geometry factors $\eta = \{0.3, 0.6 \}$. The
resulting waveforms and their key emission metrics are listed as
``Torus Fragmentation Instability'' and enumerated by Piro1--Piro4 in
Table~\ref{tab:phenowaveforms}. The full waveform names encode the
particular parameter values used. As in the case of the bar-mode
emission model, we emphasize that also the Torus Fragmentation
Instability represents an extreme GW emission scenario for CCSNe. It
may be unlikely based on theoretical considerations (e.g.,
\cite{ott:09,santamaria:11dcc}), but has not been ruled out
observationally.

\subsubsection{Ad-Hoc Waveforms: sine-Gaussians}
 
Following previous GW searches, we also employ \textit{ad-hoc} sine-Gaussian
waveforms to establish frequency-dependent upper limits on the emitted
energy in GWs. This also allows us to compare the sensitivity of our
targeted search with results from previous all-sky searches for GW
bursts (e.g., \cite{ligo_burst_s5y1:09,S5y2Burst,S6Burst})

Sine-Gaussian waveforms are, as the name implies, sinusoids in a
Gaussian envelope. They are analytic and given by
\begin{flalign}
\label{eqn:sgp}
h_+(t)&= A \frac{1+\alpha^2}{2} \exp {(-t^2/\tau^2)} \sin(2\pi f_0 t) \, , \\
\label{eqn:sgc}
h_\times(t)&= A \alpha \exp {(-t^2/\tau^2)} \cos(2\pi f_0 t) \, .
\end{flalign}
Here, $A$ is an amplitude scale factor, $\alpha=\cos\iota$ is the
ellipticity of the waveform with $\iota$ being the inclination angle,
$f_0$ is the central frequency, and $\tau=Q/(\sqrt{2}\pi f_0)$, where
$Q$ is the quality factor controlling the width of the Gaussian and
thus the duration of the signal. Since the focus of our study is more
on realistic and phenomenological waveforms, we limit the set of 
sine-Gaussian waveforms to four, enumerated SG1--SG4 in
Table~\ref{tab:phenowaveforms}. We fix $Q=8.9$ and study linearly 
polarized ($\cos\iota = 0$) and elliptically polarized ($\cos\iota$ 
sampled uniformly on $[-1,1]$) waveforms at $f = \{235,
1304\}\,\mathrm{Hz}$. We choose this quality factor and these
particular frequencies for comparison with
\cite{ligo_burst_s5y1:09,S5y2Burst,S6Burst}.

\subsection{Systematic Uncertainties}

Our efficiency estimates are subject to a number of uncertainties. 
The most important of these are calibration uncertainties in the 
strain data recorded at each detector, and Poisson uncertainties 
due to the use of a finite number of injections (Monte Carlo uncertainties).
We account for each of these uncertainties in the sensitivities reported 
in this paper.

We account for Poisson uncertainties from the finite number of injections 
using the Bayesian technique described in \cite{Paterno:2004cb}.
Specifically, given the total number of injections performed at some 
amplitude and the number detected, we compute the 90\% credible 
lower bound on the efficiency assuming a uniform prior on $[0,1]$ 
for the efficiency. All efficiency curves reported in this paper are 
therefore actually 90\% confidence level lower bounds on the efficiency.

Calibration uncertainties are handled by rescaling quoted $\hrss$ and 
distance values following the method in \cite{ligo_hfburst:09}.
The dominant effect is from the uncertainties in the amplitude calibration; 
these are estimated at approximately 10\% for G1, H1, and H2, 
14\% for L1, and 6\%-8\% for V1 at 
the times of the two CCSNe studied~\cite{Abadie2010223,marion:08}. 
The individual detector amplitude uncertainties are combined into a single uncertainty 
by calculating a combined root-sum-square signal-to-noise ratio and 
propagating the individual uncertainties assuming each error is independent 
(the signal-to-noise ratio is used as a proxy for the 
loudness measures the two pipelines use for ranking events).
This combination depends upon the relative sensitivity of each detector, 
which is a function of frequency, so we compute the total uncertainty at 
a range of frequencies across our analysis band for each CCSN and select the largest 
result, 7.6\%, as a conservative estimate of the total 1-$\sigma$ uncertainty.
This 1-$\sigma$ uncertainty is then scaled by a factor of 1.28 
(to 9.7\%) to obtain the factor by which our amplitude and distance 
limits must be rescaled in order to obtain values consistent with a 90\% 
confidence level upper limit.

\section{Search Results}\label{sec:results}

As discussed in Section~\ref{sec:overview}, on-source events from 
each combination of CCSN, detector network, and pipeline are assigned 
a false alarm rate by comparing to time-slide background events. 
Table~\ref{fapcwb3} lists the FAR values of the loudest event found 
by each pipeline for each network and CCSN. The lowest FAR, 
1.7\,$\times10^{-6}$\,Hz, was reported by {\cwb} for the analysis of 
SN~2007gr with the H1H2L1V1 network. This rate can be converted to a 
false alarm probability (FAP) using equation (\ref{eqn:fap}). 
The total duration of data processed by {\cwb} or {\xp} for the two 
CCSNe was $T_\mathrm{on} = 873461$\,s. Equation (\ref{eqn:fap}) then 
yields a false-alarm probability of 0.77 for the loudest event; this is 
consistent with the event being due to background noise. 
We conclude 
that none of the events has a FAP low enough to be considered as a 
candidate GW detection.

\begin{table}[htb]
\label{fapcwb3}
\centering
\caption{False alarm rate (FAR) of the loudest event found by each pipeline 
for each detector network. No on-source events survived the coherent tests 
and data quality cuts for the {\cwb} analysis of the H1H2L1 and H1H2 networks 
for SN~2007gr. The lowest FAR, 1.7\,$\times10^{-6}$\,Hz, corresponds to a FAP of 0.77.}
\begin{tabular}{l@{\hspace*{1em}}l@{\hspace*{1em}}l}
\hline
\hline
\multicolumn{1}{c}{Network} & \multicolumn{1}{c}{{\cwb}} & \multicolumn{1}{c}{{\xp}} \\
\hline
  H1H2L1V1 & 1.7\,$\times10^{-6}$\,Hz  &  2.5\,$\times10^{-6}$\,Hz \\
  H1H2L1   & no events                 &  1.1\,$\times10^{-5}$\,Hz \\
  H1H2V1   & 1.2\,$\times10^{-5}$\,Hz  &  5.3\,$\times10^{-6}$\,Hz \\
  H1H2     & no events                 &  7.1\,$\times10^{-5}$\,Hz \\
  L1V1     & 4.8\,$\times10^{-5}$\,Hz  &  4.1\,$\times10^{-3}$\,Hz \\ 
  G1V1     & 1.2\,$\times10^{-5}$\,Hz  &  2.7\,$\times10^{-5}$\,Hz \\ \hline  
\hline
\end{tabular}
\end{table}

We note that the loudest events reported by {\cwb} and {\xp} are both from 
the analysis of SN~2007gr with the H1H2L1V1 network; this is consistent with 
chance as this network combination accounted for more than 60\% of the data 
processed. In addition, the times of the loudest X-pipeline and cWB events differ by more than a day, 
so they are not due to a common physical trigger.

\subsection{Detection efficiency vs.~distance}

Given the loudest event, we can compute detection efficiencies for the 
search following the procedure detailed in Section~\ref{sec:simulations}. 
In brief, we measure the fraction of simulated signals that produce 
events surviving the coherent tests and data quality cuts and which have 
a FAR (or equivalently FAP) lower than the loudest event.

\begin{figure*}[t] 
  \begin{minipage}[c][][t]{0.495\textwidth}
    \vspace*{\fill}
    \flushleft
    \includegraphics[width=0.96\linewidth]{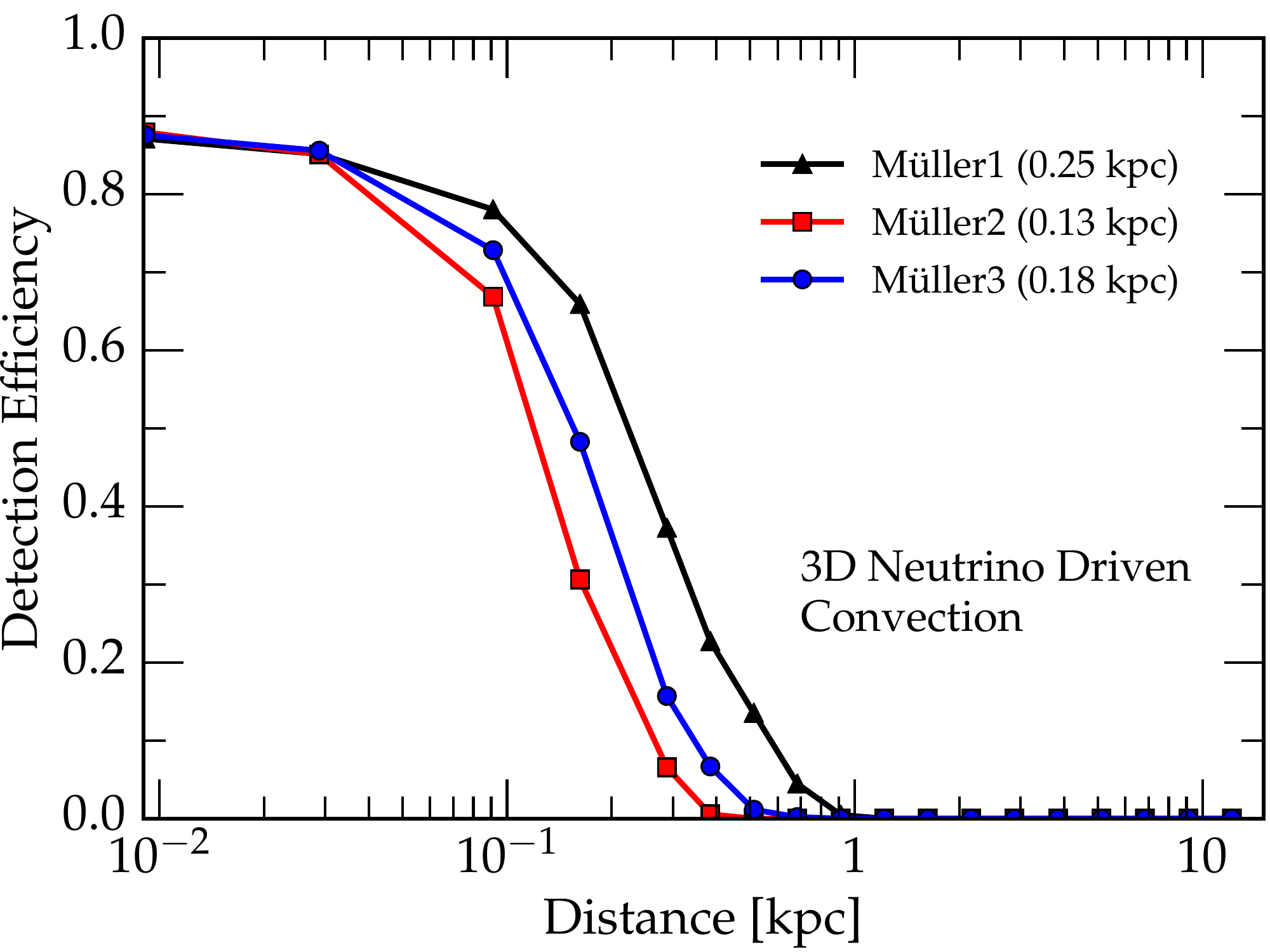}
    \includegraphics[width=0.96\linewidth]{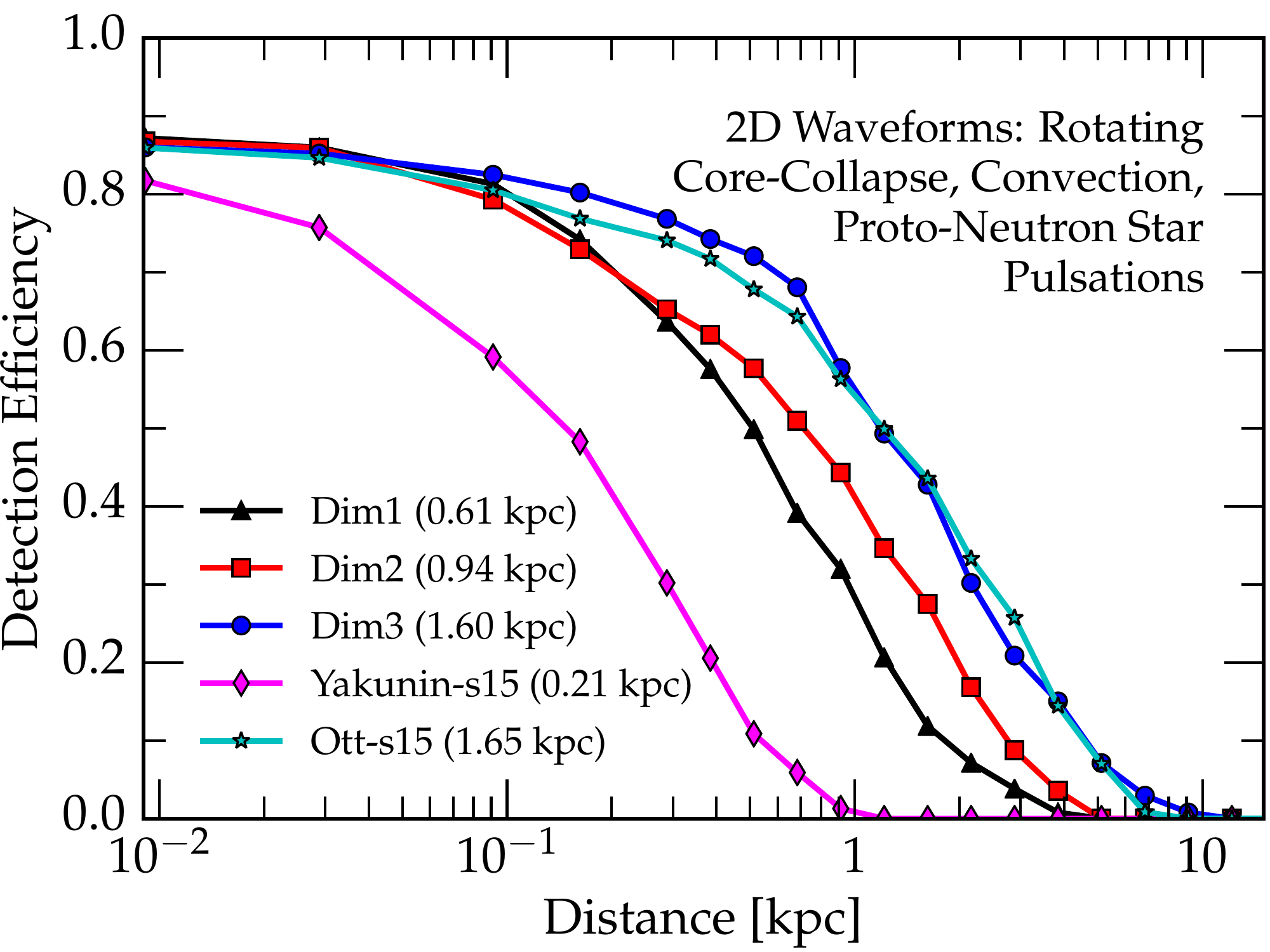}
  \end{minipage}
  \begin{minipage}[c][][t]{0.495\textwidth}
    \vspace*{\fill}
    \flushright
    \includegraphics[width=0.96\linewidth]{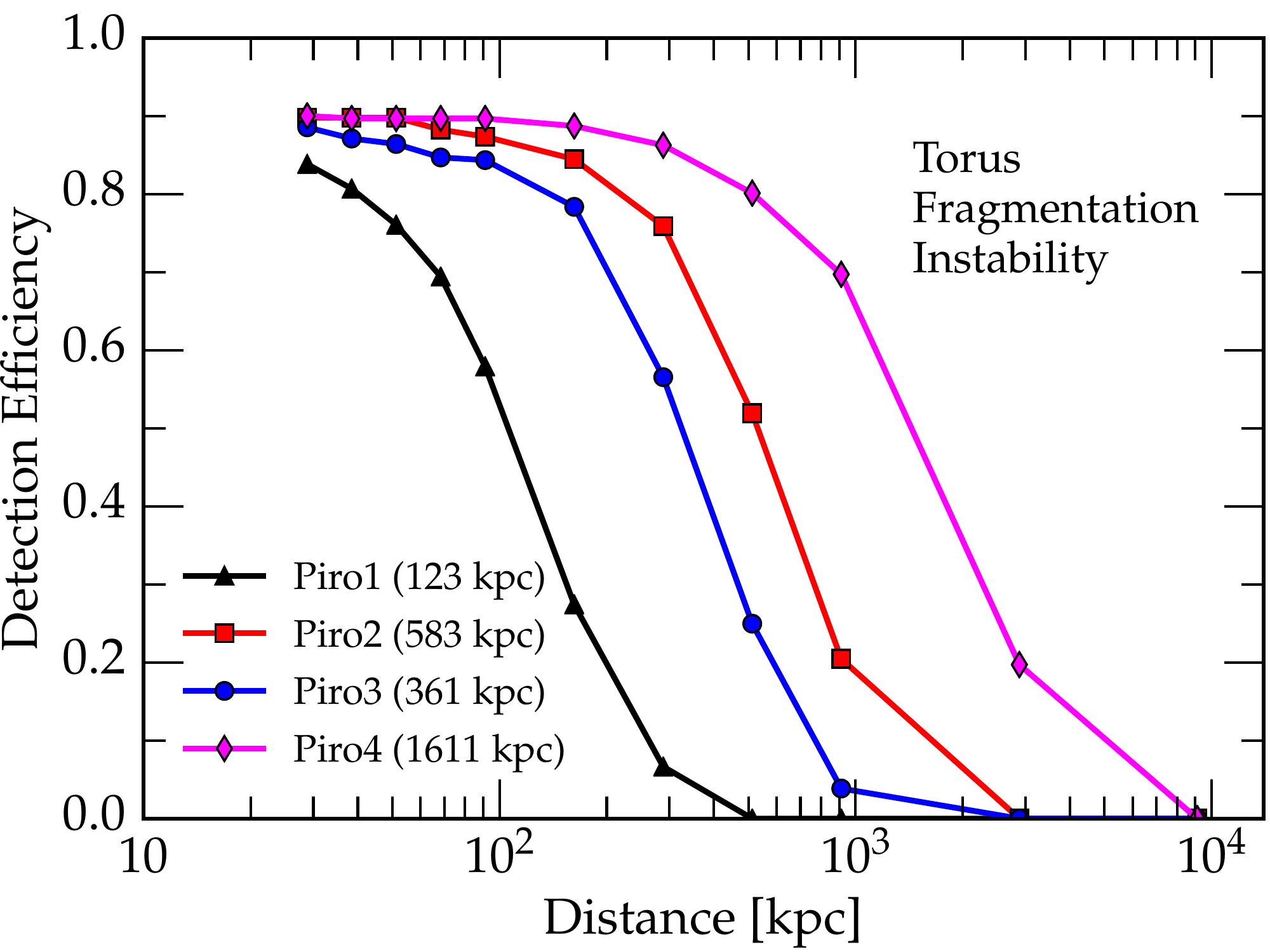}
    \includegraphics[width=0.96\linewidth]{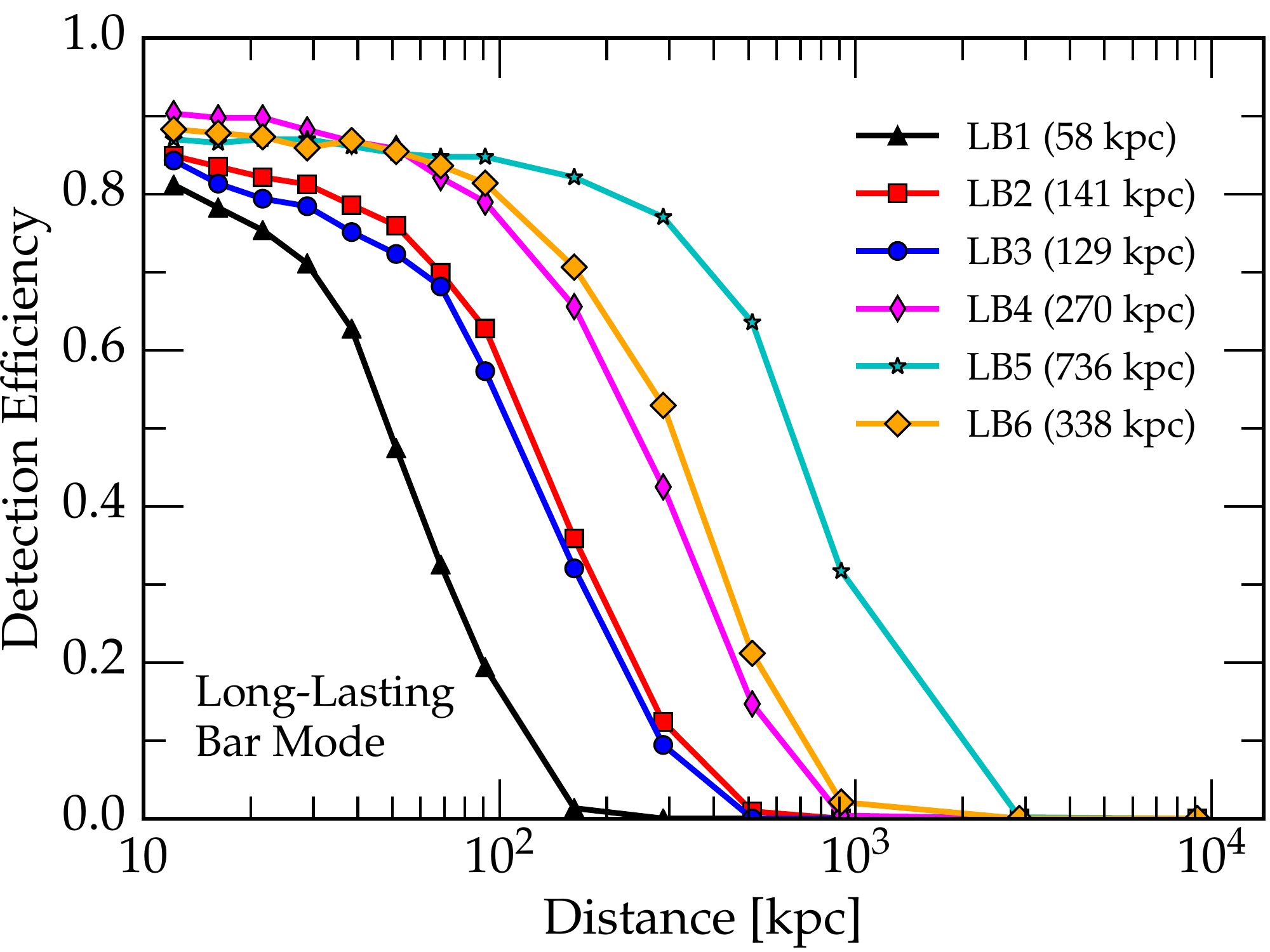}
  \end{minipage}
  \caption{\label{fig:sn2007gr}
          SN~2007gr detection efficiency versus distance for the 
		  waveforms from multi-dimensional CCSN simulations (left) and the phenomenological waveforms (right) 
		  described in Tables~\ref{tab:numwaveforms} and \ref{tab:phenowaveforms}.
		  Simulated GW signals are added into detector data with a range of 
		  amplitudes corresponding to different source distances. A simulated signal 
		  is considered detected if {\cwb} or {\xp} reports an event that survives 
		  the coherent tests and data quality cuts with a FAR value lower than 
		  that of the loudest event from the SN~2007gr and SN~2011dh on-source windows. 
		  These efficiencies are averaged over all detector network combinations for SN~2007gr.
		  The efficiencies are limited to $\le93\%$ at small distances due to the fact that 
		  this was the duty cycle for coincident observation over the SN~2007gr on-source window.
		  The numbers in brackets for each model are the distances at which the efficiency equals $50\%$ 
of the asymptotic value at small distances.  }
\end{figure*}

\begin{figure*}[t]
  \begin{minipage}[c][][t]{0.495\textwidth}
    \vspace*{\fill}
    \flushleft
    \includegraphics[width=0.96\linewidth]{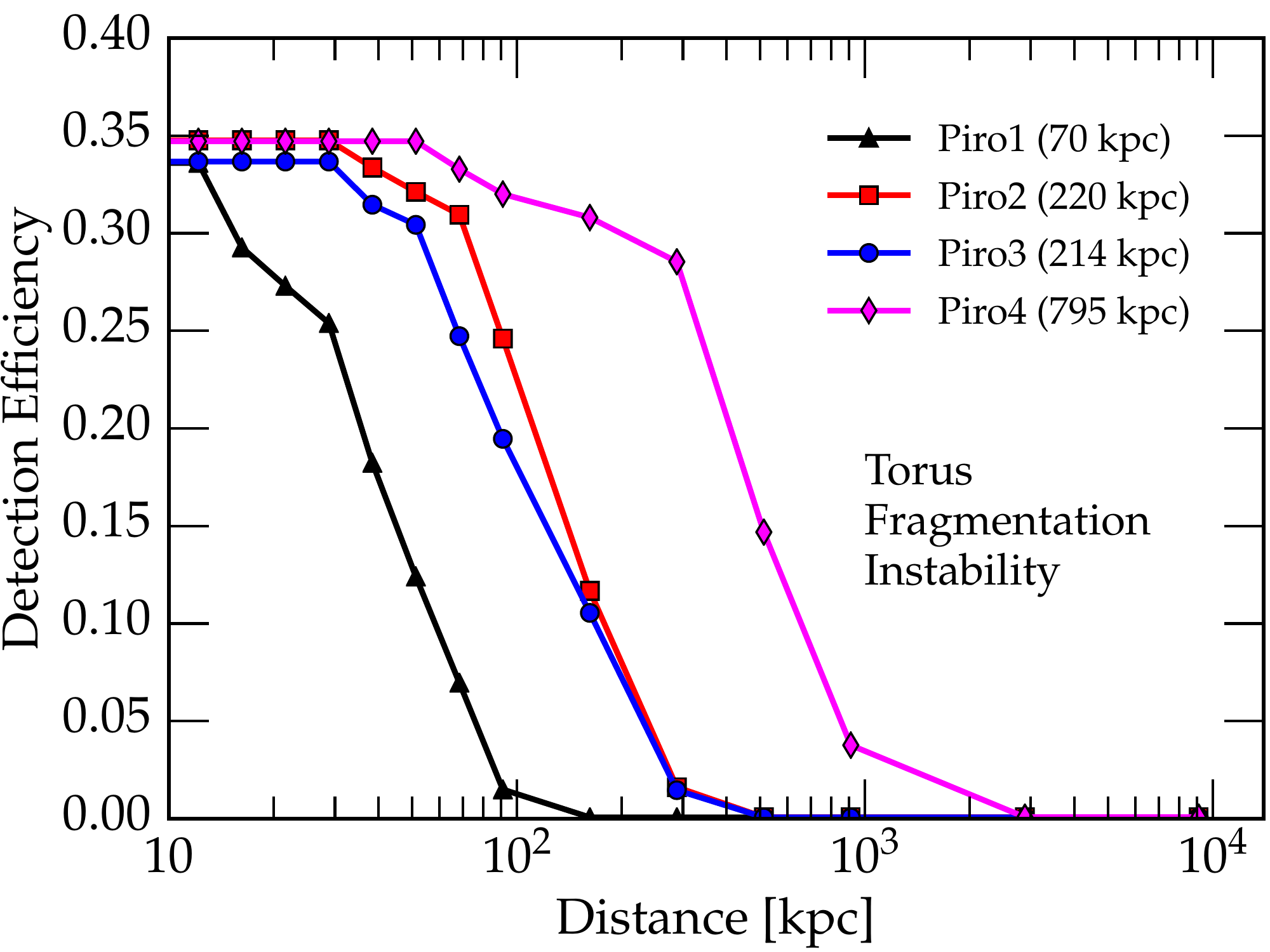}
  \end{minipage}
  \begin{minipage}[c][][t]{0.495\textwidth}
    \vspace*{\fill}
    \flushright
    \includegraphics[width=0.96\linewidth]{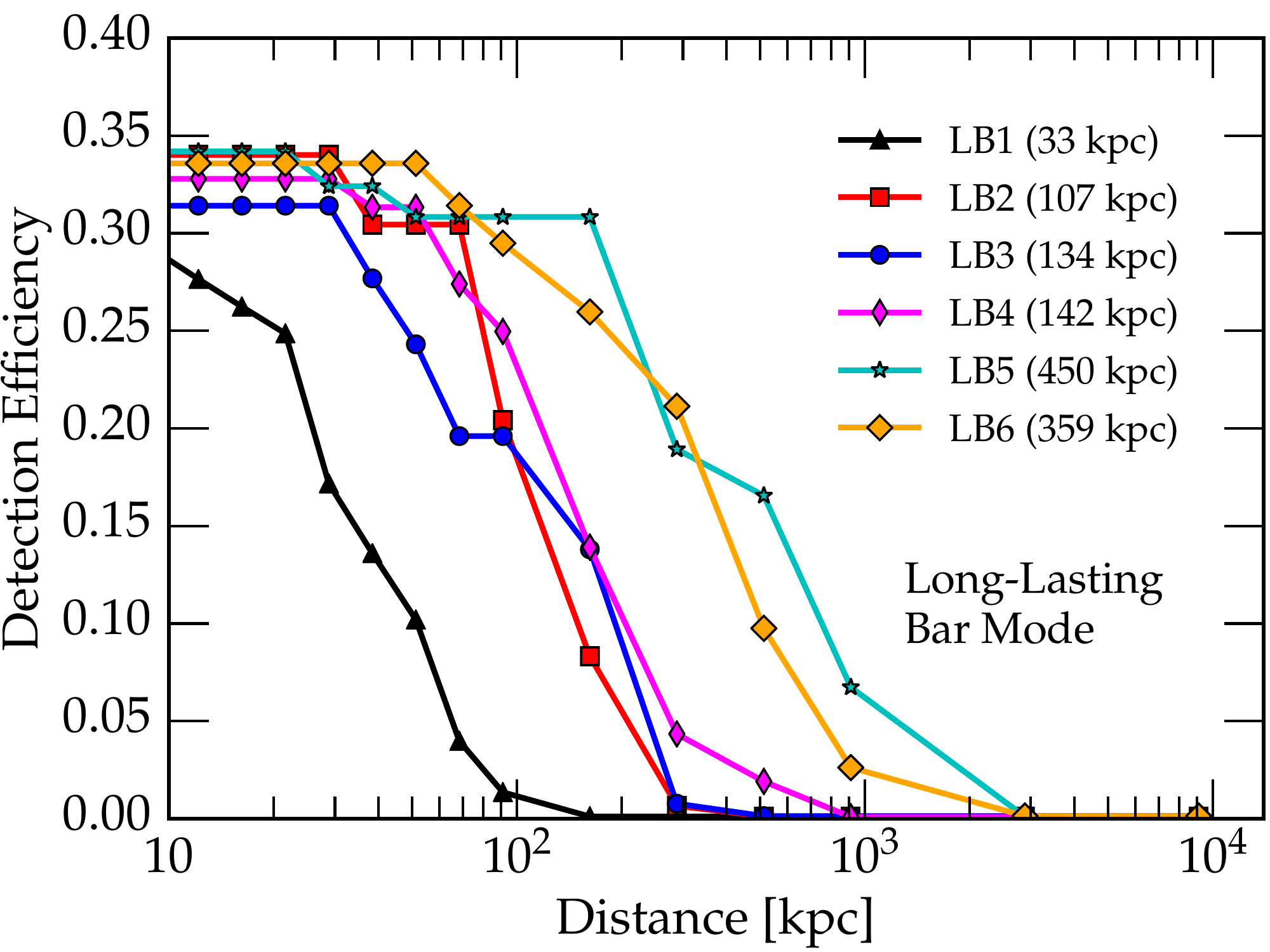}
  \end{minipage}
  \caption{\label{fig:sn2011dh}
          SN~2011dh detection efficiency versus distance for the phenomenological waveforms 
		  described in Table~\ref{tab:phenowaveforms}. 
		  Simulated GW signals are added into detector data with a range of 
		  amplitudes corresponding to different source distances. A simulated signal 
		  is considered detected if either {\cwb} or {\xp} reports an event that survives 
		  the coherent tests and data quality cuts with a FAR value lower than 
		  that of the loudest event from the SN~2007gr and SN~2011dh on-source windows. 
		  The efficiencies are limited to $\le37\%$ at small distances due to the fact 
		  that this was the duty cycle for coincident observation over the SN~2011dh 
		  on-source window; some simulations are also vetoed by data quality cuts.
		  The numbers in the brackets are the distances at which the efficiency 
		  equals $50\%$ of its maximum value for each model.  }
\end{figure*}

Figures~\ref{fig:sn2007gr} and \ref{fig:sn2011dh} show the efficiency
as a function of distance for the CCSN waveforms from
multi-dimensional simulations and the phenomenological waveforms
discussed in Section \ref{sec:simulations} and summarized in
Tables~\ref{tab:numwaveforms} and \ref{tab:phenowaveforms}.  For
SN~2007gr, the maximum distance reach is of order 1\,kpc for waveforms
from detailed multi-dimensional CCSN simulations, and from $\sim$100\,kpc to
$\sim$1\,Mpc for GWs from the phenomenological models (torus fragmentation
instability and long-lived rotating bar mode). The variation in
distance reach is due to the different peak emission frequencies of
the models and the variation in detector sensitivities with frequency
as shown in Figure~\ref{fig:noise}.  The distance reaches for
SN~2011dh are lower by a factor of several than those for SN~2007gr;
this is due to the difference in sensitivity of the operating
detectors, as also evident in Figure~\ref{fig:noise}.  Finally, we
note that at small distances the efficiencies asymptote to the
fraction of the on-source window that is covered by coincident data,
approximately 93\% for SN~2007gr and 37\% for SN~2011dh (up to a few
percent of simulated signals are lost due to random coincidence with
data quality cuts).

We do not show the efficiencies for the multi-dimensional simulation CCSN waveforms for SN~2011dh, 
as the detection efficiency was negligible in this case. This is due to the fact that 
the relative orientation of the G1 and V1 detectors -- rotated approximately $45^\circ$ 
with respect to each other -- means that the two detectors are sensitive to orthogonal 
GW polarizations. In order for the coherent cuts to reject background noise 
{\xp} needs to assume some relationship between these two polarizations.
We require that the $h_+$ and $h_\times$ polarizations are out of phase by $90^\circ$, 
as would be expected for emission from a rotating body with a non-axisymmetric 
quadrupole deformation. We choose this because the strongest GW emission models 
are for rotating non-axisymmetric systems (the fragmentation instability and long-lived bar mode). 
Unfortunately, the waveforms from multi-dimensional CCSN simulations are either 
linearly polarized (i.e.~have only one polarization) or exhibit 
randomly changing phase. Hence, they cannot be detected by the 
search performed for SN~2011dh with {\xp}.
The tuning of {\cwb} did not use these constraints, however the G1 noise floor was 
about a factor of 2 higher than V1 around 1000\,Hz and the difference was even greater at lower frequencies.
This issue weakened the internal {\cwb} measures of correlation of the reconstructed 
signal between the two interferometers and severely reduced the detection efficiencies 
at distances beyond a few parsecs. 

The distances shown in Figures~\ref{fig:sn2007gr} and \ref{fig:sn2011dh} 
show the probability of a GW signal producing an event with FAP lower than 
that of the loudest event.
The physical interpretation of the efficiency $\epsilon$ at a distance $d$ for a 
given model is related to the prospect of excluding the model with observations.
Explicitly, the non-observation of any events with FAP lower than the 
loudest event gives a frequentist exclusion of that GW emission model for a 
source at distance $d$ with confidence $\epsilon$.
However, in this search the loudest event had a large FAP (0.77). 
In order for an event to be considered as a possible detection it would need to 
have a FAP of order $10^{-3}$ or less; we find that imposing this more stringent 
requirement lowers the maximum distance reach by approximately 5\%-25\% 
depending on the waveform model.

Unfortunately, none of the models have distance reaches out to the $\sim$10\,Mpc distance  
of SN~2007gr or SN~2011dh; we conclude that our search is not able to constrain
the GW emission model for either of these CCSNe.

\subsection{Constraints on Energy Emission}

In addition to the astrophysically motivated phenomenological and 
multi-dimensional CCSN simulation waveforms, we employ the \textit{ad-hoc} 
sine-Gaussian waveforms specified by equations (\ref{eqn:sgp}) and (\ref{eqn:sgc})
to establish frequency-dependent upper limits on the emitted energy in GWs. 
This also allows us to compare the sensitivity of our
targeted search with results from previous all-sky searches for GW
bursts (e.g., \cite{ligo_burst_s5y1:09,S5y2Burst,S6Burst})

The detection efficiency is computed using the same procedure as for
the other waveforms. However, since these \textit{ad hoc} waveforms
have no intrinsic distance scale, we measure the efficiency as a
function of the root-sum-square amplitude $\hrss$, defined by
equation~(\ref{eq:hrss}).  For this study, we use the two
sine-Gaussian waveforms described in Section~\ref{sec:simulations},
which have central frequencies of 235\,Hz and 1304\,Hz. These are
standard choices for all-sky burst searches~\cite{S6Burst}.
Table~\ref{tab:energy} lists the $\hrss$ values at which the
efficiency reaches half of its maximum value. Note that we use the
half-maximum efficiency rather than 50\% efficiency here, since the
maximum efficiency is limited by the fraction of the on-source window
that is covered by coincident data. The half-maximum gives a measure
of the distance reach of the instruments independent of their duty cycle.

These $\hrss$ values can be converted to limits on energy emission by
assuming a specific angular emission pattern of the source
\cite{sutton:13}.  For simplicity, we assume isotropic emission, for
which
\begin{equation}
\label{eqn:energy}
E_\mathrm{GW} = \frac{\pi^2 c^3}{G}D^2 f^2_0\hrss^2 \, .
\end{equation}
Here $f_0$ is the peak frequency of the GW and $D$ is the distance of
the source. We use distances of 10.55\,Mpc for SN 2007gr and 8.40\,Mpc
for SN 2011dh. Table~\ref{tab:energy} also lists the energy emission
values at which the efficiency reaches half of its maximum value. If
the total amount of energy emitted in GWs was larger than the numbers
quoted in the Table, we would have had a greater than $50\%$ chance of
seeing a signal from the CCSN at the estimated distance, provided
coincident observation with the most sensitive detector network. Note,
however, that the on-source window did not have $100\%$ coverage (see
Section~\ref{sec:networks}). 

The most stringent constraints are a few percent of a mass-energy
equivalent of a solar mass emitted in GWs at 235\,Hz, where the noise
floor is low.  The 1304\,Hz results indicate that with this data
set, we should not expect to be able to detect extra-Galactic GWs at
kHz frequencies, since the limits are less stringent, $O(10)$~$M_\odot
c^2$ or more.

The above results can be compared with the energy available in CCSNe,
which are powered by the gravitational energy released in core
collapse. The total available energy is set by the binding energy of a
typical $1.4\,M_\odot$ neutron star and is roughly
$3\times10^{53}\,\mathrm{erg}$, corresponding to $\sim$$0.15\,M_\odot
c^2$ (e.g., \cite{lattimer:01}). The observation of neutrinos from
SN~1987A confirmed that $\sim$$99\%$ of that energy is emitted in the
form of neutrinos in proto-neutron star cooling (e.g.,
\cite{vissani:15}). The typical CCSN explosion kinetic energy is
$\sim$$10^{51}\,\mathrm{erg}$ ($\sim$$10^{-3}\,M_\odot
c^2$). Considering these observational constraints, the energy emitted
in GWs is unlikely to exceed $O(10^{-3})\,M_\odot c^2$. Hence, the energy
constraints obtained by this search for SNe 2007gr and 2011dh are not
astrophysically interesting.

\begin{table*}
\caption{Gravitational-wave energy emission constraints at
  half-maximum detection efficiency for SN~2007gr and SN~2011dh.
  These assume distances of 10.55\,Mpc for SN 2007gr and 8.40\,Mpc for
  SN 2011dh.
\label{tab:energy}}
\begin{center}
\begin{tabular}{r|c|c|c|c|c|c}
\hline
\hline 
\multirow{2}{*}{Waveform~~}& \multicolumn{3}{|c|}{SN~2007gr} & \multicolumn{3}{|c}{SN~2011dh} \\ \cline{2-7}
   				  &  $\hrss$\,[\hrssu]  &  $E_\mathrm{GW}\,[\mathrm{erg}]$  &  $E_\mathrm{GW}\,[M_{\odot}c^2]$ 
				  &  $\hrss$\,[\hrssu]  &  $E_\mathrm{GW}\,[\mathrm{erg}]$  &  $E_\mathrm{GW}\,[M_{\odot}c^2]$ \\ \hline
SGel2  SG235Q9    & $5.4\times10^{-22}$ & 6.7$\times10^{52}$ &   0.038   & $9.1\times10^{-21}$ & 1.2$\times10^{55}$ &  6.8        \\ \hline
SGlin2 SG235Q9    & $6.6\times10^{-22}$ & 1.0$\times10^{53}$ &   0.058   & $4.8\times10^{-20}$ & 3.4$\times10^{56}$ &  1.9$\times10^2$      \\ \hline
SGel2  SG1304Q9   & $2.1\times10^{-21}$ & 3.1$\times10^{55}$ &  17       & $2.2\times10^{-21}$ & 2.3$\times10^{55}$ &  13       \\ \hline
SGlin2 SG1304Q9   & $2.5\times10^{-21}$ & 4.6$\times10^{55}$ &  26       & n/a               & n/a                &  n/a     \\ \hline
\hline
\end{tabular} \label{table:res}
\end{center}
\end{table*}

\subsection{Model Exclusion Confidence}

As we have seen, it is unlikely that we will have coincident science-quality 
data covering an entire multi-day on-source window for any given CCSN. 
In the present analysis, the coverage of the on-source windows is 
approximately 93\% for SN~2007gr and 37\% for SN~2011dh. Considering that 
data-quality cuts typically remove another few percent of livetime, we 
cannot expect to exclude even fairly strong GW emission at the 90\% confidence level 
for a single CCSN. However, by combining observations of multiple CCSNe, it 
is straightforward to exclude the simple model in which all CCSNe produce 
identical GW signals; i.e., assuming \textit{standard-candle} emission.

Consider a CCSN model $M_{\mathrm{SN}}$ which predicts a particular GW 
emission pattern during the CCSN event (e.g., one of the waveforms 
considered in Section~\ref{sec:simulations}).
In the case that no GW candidates are observed, we can 
constrain that model using observations from multiple CCSN events at 
known distances $d_i$ using the measured detection efficiencies 
$\epsilon_i(d_i)$ for each supernova (e.g., as in Figure~\ref{fig:sn2007gr}).  
These $\epsilon_i(d_i)$  can be combined into an overall model exclusion 
probability~\cite{kalmus:13a}, $P_\mathrm{excl}$:
\be
P_\mathrm{excl} = 1 - \prod_{i =1}^N (1-\epsilon_i(d_i)) \label{eq:reach}
\ee 
It is also straightforward to marginalize over uncertainties in the $d_i$ 
(as in Table~\ref{table:triggers}) by the replacement
\begin{equation}
\label{eqn:marginalise}
\epsilon_i(d_i) \to \epsilon_i \equiv \int_0^\infty \!\!\!d\bar{d} \,\pi_i(\bar{d}) \epsilon_i(\bar{d})
\end{equation}
where $\pi_i$ is our prior on the distance to CCSN $i$ (e.g., a Gaussian).

In the light of the measured sensitivity ranges in
Figures~\ref{fig:sn2007gr} and \ref{fig:sn2011dh}, it is clear that we
cannot exclude any of the considered models of GW emission for
SN~2007gr and SN~2011dh with the current data.  However, iLIGO and
Virgo are being upgraded to advanced configurations, with a final
design sensitivity approximately a factor of ten better than for the
period 2005-2011 considered in this paper. It is therefore instructive
to consider what model exclusion statements the advanced detectors
will be able to make using future CCSNe similar to SN~2007gr and
SN~2011dh.  We will focus on the phenomenological waveform models of plausible but more
extreme GW emission, where we expect to reach sooner large standard candle
model exclusion probabilities.
Specifically, we analyze the rotating bar and torus fragmentation scenarios
(see also the discussion in \cite{gossan:16}).

\begin{figure*}[t]
  \begin{minipage}[c][][t]{0.495\textwidth}
    \vspace*{\fill}
    \flushleft
    \includegraphics[width=0.96\linewidth]{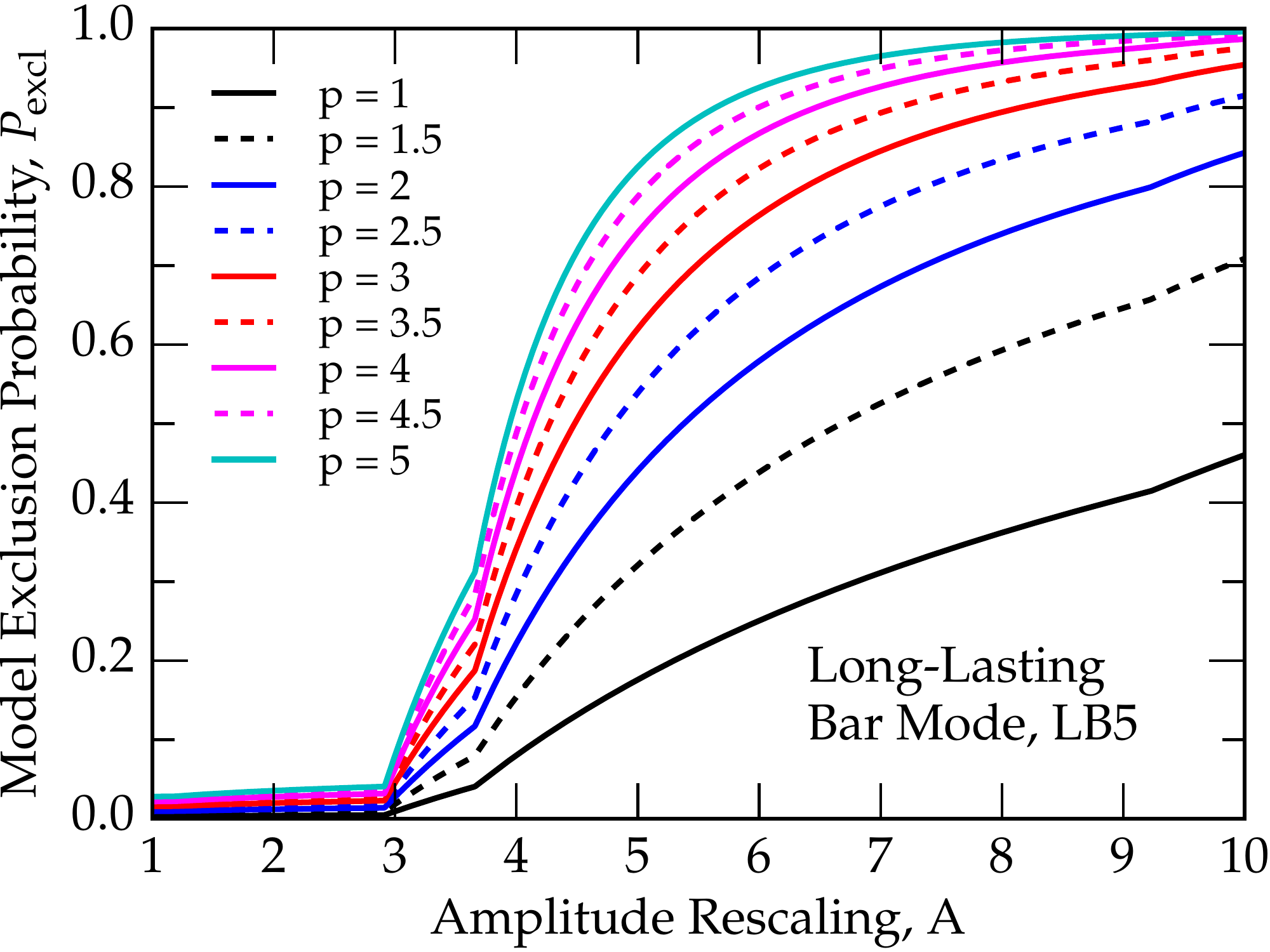}
    \includegraphics[width=0.96\linewidth]{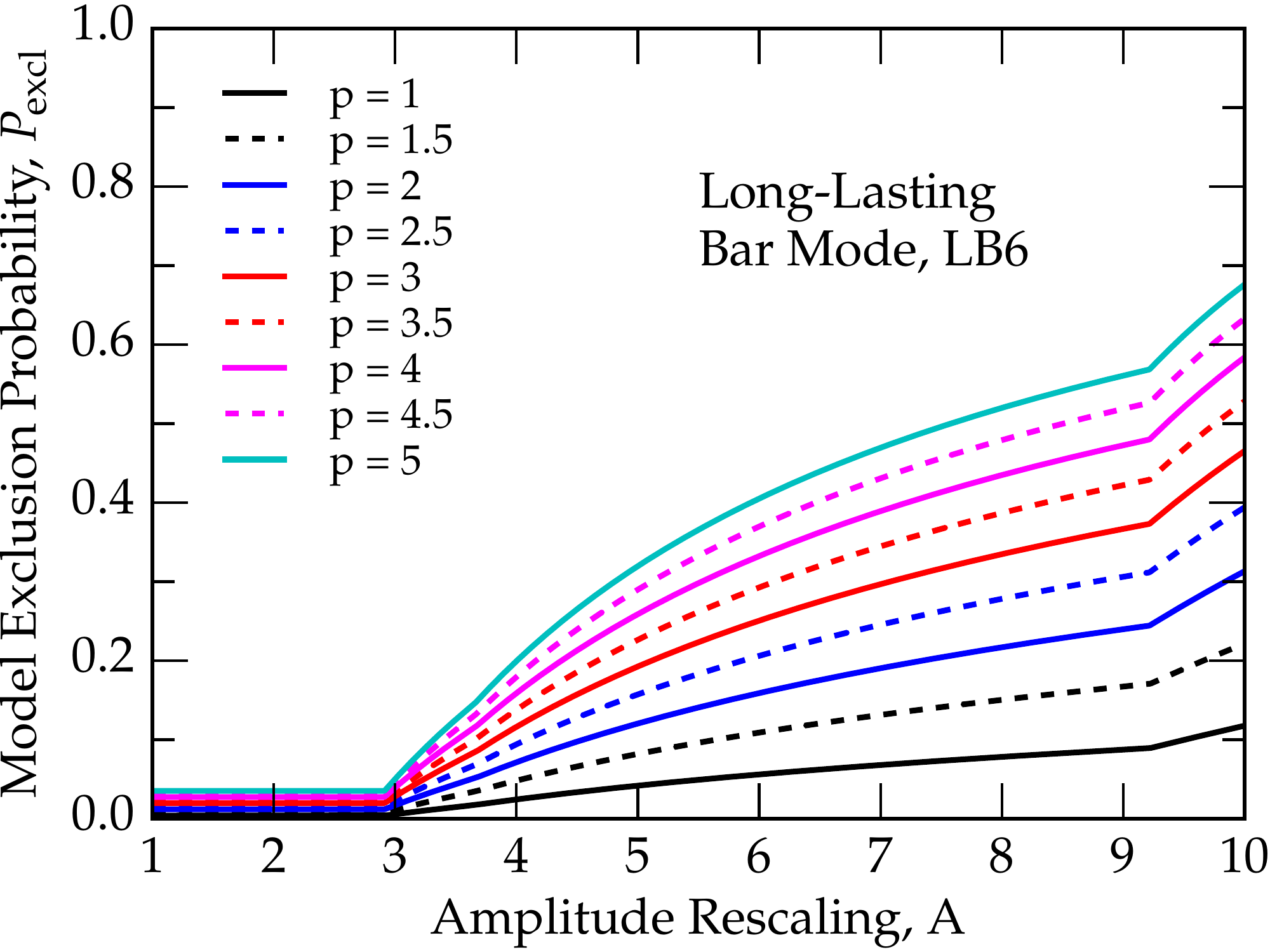}
  \end{minipage}
  \begin{minipage}[c][][t]{0.495\textwidth}
    \vspace*{\fill}
    \flushright
    \includegraphics[width=0.96\linewidth]{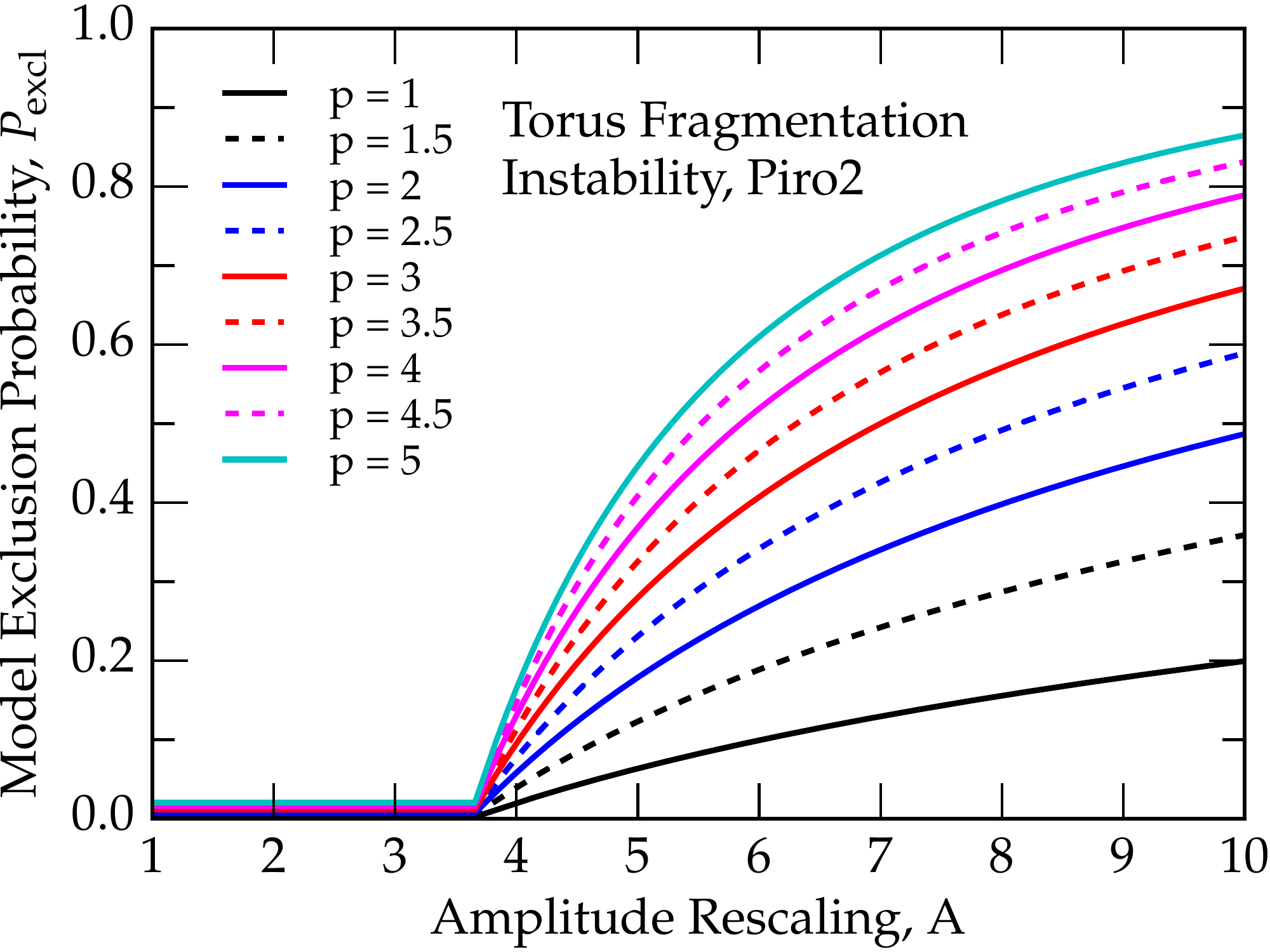}
    \includegraphics[width=0.96\linewidth]{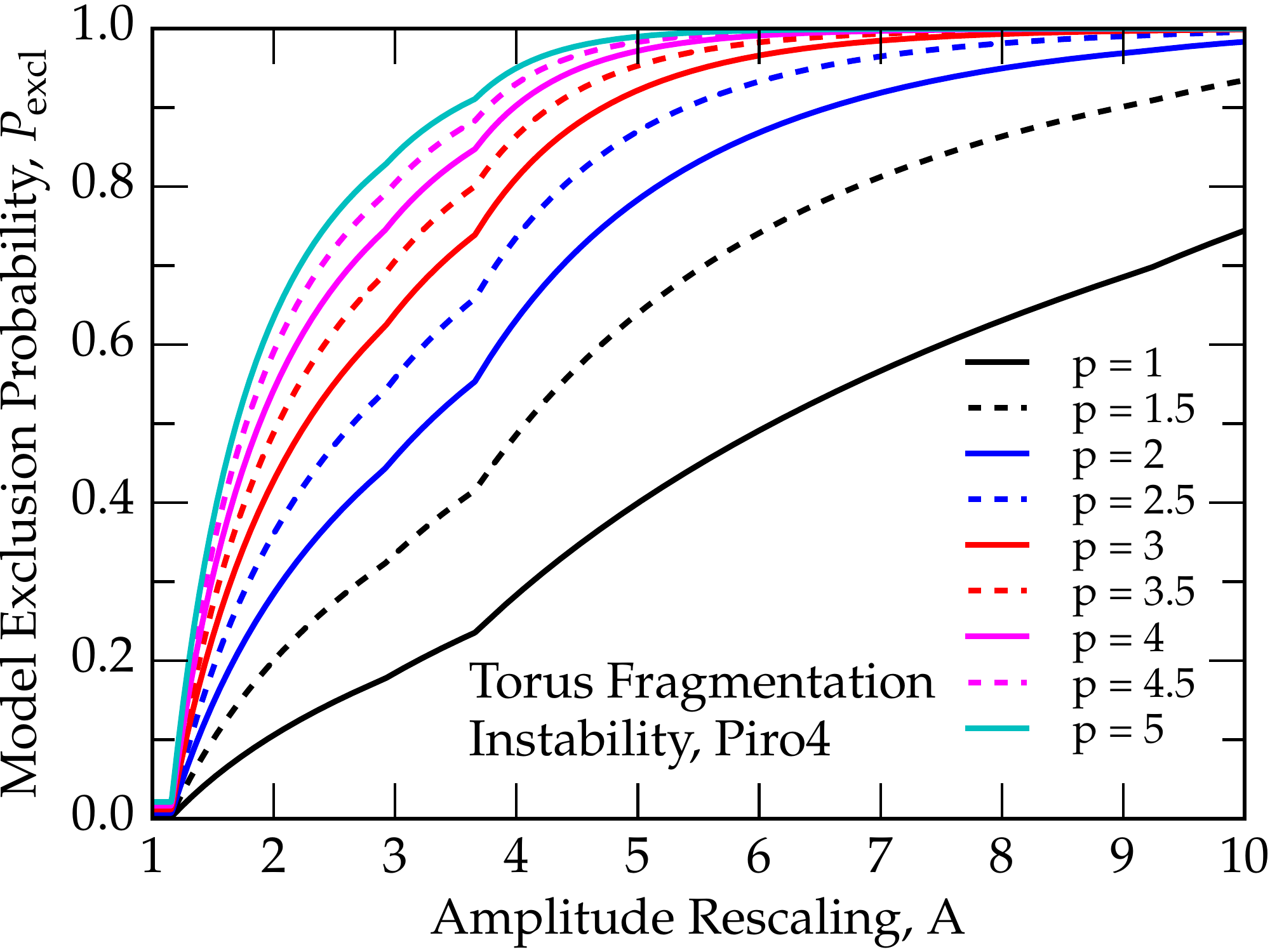}
  \end{minipage}
  \caption{\label{fig:exclusion}
  Expected model exclusion probabilities for example waveforms as a function 
  of amplitude sensitivity rescaling, $A$, and supernova sample 
  size rescaling, $p$, based on the SN~2007gr and SN~2011dh sample
  (e.g. $p=5$ corresponds to 10 supernovae). 
  The naming convention is described in Table \ref{tab:phenowaveforms}.
  Currently none of the emission models can be exluded,
  but for the advanced detectors with better sensitivity and
  more nearby CCSNe it is realistic to expect to
  rule out some of the extreme emission models.
  }
\end{figure*}

Figure~\ref{fig:exclusion} presents model exclusion confidence plots
for four of the phenomenological waveform models. These plots are
based on the measured efficiencies for SN~2007gr and SN~2011dh, but
assume the detector noise spectra have been lowered by a factor of
$A$, so the search would be expected to have the same efficiency for a
particular source at $A$-times the distance, and the number of CCSNe
in the sample has been increased by a factor of $p$.  For example,
$A=10$ represents having a sensitivity 10 times better than the
2005--2011 data, which is realistic for Advanced LIGO and Advanced
Virgo, while $p=2$ corresponds to having two CCSNe similar to
SN~2007gr and two similar to SN~2011dh.  The curves correspond to the
experimentally derived values based on the 2005-2011 data set.  It is
worth stressing that the power of excluding models from this data set
almost exclusively originates from SN~2007gr, given the more sensitive
interferometers available at the time of that supernova.  For example,
in the bottom left panel of Fig.~\ref{fig:exclusion}, the curves, when
A is smaller than 9, depend almost exclusively on the contribution of
SN~2007gr. In this regard, the presented model exclusion probabilities
will be reached with less than $2p$ CCSNe if the sample contains more
data sets comparable in coverage and sensitivity to the rescaled
SN~2007gr data set than a rescaled SN~2011dh data set. In summary,
Fig.~\ref{fig:exclusion} shows that it is a reasonable expectation
that extended coincident observations with advanced-generation
detectors will rule out extreme CCSN emission models.

\subsection{Sensitivity Advantage of the Triggered Search}

As noted in Section~\ref{sec:introduction}, targeted searches have the 
advantage over all-time all-sky searches that potential signal candidates in the 
data streams have to arrive in a well-defined temporal on-source window 
and have to be consistent with coming from the sky location of the source. 
Both constraints can significantly reduce the noise background. 
Here we assess the improved sensitivity of a triggered search 
by comparing our $\hrss$ sensitivities to linearly polarized sine-Gaussian 
waveforms for SN~2007gr to those of an all-sky search of the same data.

The most straightforward way to compare two searches is to fix the FAR  
threshold and compare the $\hrss$ values at 50\% efficiency. 
The S5/VSR1 all-sky all-time search \cite{S5y2Burst} using {\cwb} was run 
on 68.2 days of coincident H1H2L1V1 data with thresholds to give 
a FAP of 0.1 or less in the frequency band up to 2000\,Hz. 
The livetime for the {\cwb} SN~2007gr analysis of the H1H2L1V1 network
was 3.25 days, 
so a FAP of 0.1 corresponds to a FAR of $3.56\times10^{-7}$\,Hz. 
Including calibration and Monte Carlo uncertainties, 
the ${\hrss}$ values at 50\% efficiency for this FAR are 
$5.0\times10^{-22}\,\mathrm{Hz}^{-1/2}$ at 235\,Hz and 
$2.2\times10^{-21}\,\mathrm{Hz}^{-1/2}$ at 1304\,Hz. 
After adjusting for systematic differences in the antenna responses
and noise spectra\footnote{In particular, during the on-source window of SN~2007gr the noise spectral 
density for L1 was about 50 percent worse at low frequencies than the average value
during the whole of S5.} between the S5/VSR1 all-sky search and the SN~2007gr 
search, the effective all-sky ${\hrss}$ values are 
$7.0\times10^{-22}\,\mathrm{Hz}^{-1/2}$ at 235\,Hz and
$2.9\times10^{-21}\,\mathrm{Hz}^{-1/2}$ at 1304\,Hz, 
approximately 30\% to 40\% higher than the targeted search.
Equivalently, the distance reach of our targeted search is larger 
than that of the all-time all-sky search by 30\% to 40\% at this FAP. 

Alternatively, we can compare the two searches without adjusting to 
a common FAR. After allowing for systematic differences in the antenna 
responses and noise spectra between the S5/VSR1 all-sky search and the 
SN~2007gr search, we find that the ${\hrss}$ values at 50\% efficiency 
are identical (to within a few percent).
However, the FAR of the SN~2007gr search is lower by an order of magnitude: 
$1.8\times10^{-9}\,$Hz compared to $1.7\times10^{-8}\,$Hz for the all-sky search.
This is consistent with expectations for restricting from an all-sky 
search to a single sky-position search.
Furthermore, the FAP for a trigger produced by the SN~2007gr search will
be smaller than that of a trigger from the all-sky search at the same FAR
because the SN~2007gr on-source window (3.5 days for {\cwb} and {\xp} combined) is a factor of 20 shorter
than the all-sky window (68.2 days). 
So if we consider a surviving trigger 
that is just above threshold in the two searches, the SN~2007gr trigger 
will have an FAP a factor of approximately 200 lower than an all-sky trigger 
with the same $\hrss$.

\section{Summary and Discussion}
\label{sec:summaries}

We presented the results of the first iLIGO-GEO-Virgo search for
gravitational-wave (GW) transients in coincidence with optically
detected core-collapse supernovae (CCSNe) observed between 2007 and
2011. Two CCSNe, SN~2007gr and SN~2011dh, satisfied our criteria of
proximity, well-constrained time of core collapse, and occurrence
during times of coincident high-sensitivity operation of at least two
GW detectors.  No statistically significant GW events were observed
associated with either CCSN.  

We quantified the sensitivity of the search as a function of distance
to the CCSNe using both representative waveforms from detailed
multi-dimensional CCSN simulations and from semi-analytic
phenomenological models of plausible but extreme emission
scenarios. The distances out to which we find signals detectable for
SNe 2007gr and 2011dh range from $O(\lesssim 1)$\,kpc for waveforms from
detailed simulations to $O(1)$\,Mpc for the more extreme phenomenological models.  From
the known distances of our two target supernovae, we estimated the
minimum energy in gravitational waves corresponding to our sensitivity
limits using \textit{ad-hoc} sine-Gaussian waveforms. These range from
$O(0.1)\,M_\odot c^2$ at low frequencies to $\gtrsim O(10)\,M_\odot
c^2$ above 1\,kHz. 

This first search for GWs from extragalactic CCSNe places the 
most stringent observational constraints to-date on GW emission in
core-collapse supernovae. A comparison of our search's sensitivity
with the standard all-sky, all-time search for generic GW bursts in
the same GW detector data shows a 35\%-40\% improvement in distance
reach at fixed false alarm probability. This improvement comes from
knowledge of the sky positions of the CCSNe and approximate
knowledge of the collapse times. It is, hence, clearly beneficial to carry 
out targeted searches for GWs from CCSNe.

The results of our search do not allow us to constrain astrophysically
meaningful GW emission scenarios. 
We have extrapolated our results to the
sensitivity level expected for Advanced LIGO and Virgo. 
Considering the improved detector sensitivity and assuming the analysis of 
multiple CCSNe, we find that at design sensitivity (c.~2019, \cite{Aasi:2013wya}) 
this network will be able to constrain
the extreme phenomenological emission models for extragalactic CCSNe
observed out to distances of $\sim$10\,Mpc. Detection of the most realistic GW
signals predicted by multi-dimensional CCSN simulations will require a
Galactic event even at the design sensitivity of the Advanced
detectors. These are consistent with the results of the study in \cite{gossan:16}, 
which used data from iLIGO and Virgo recoloured to match the 
advanced detector design sensitivities.  We conclude that
third-generation detectors with a sensitivity improvement
of a factor of $10-20$ over the Advanced detectors may be needed to
observe GWs from extragalactic CCSNe occurring at a rate of $1-2$ per
year within $\sim$10\,Mpc.

\begin{acknowledgments}\label{sec:acknowledgments}

We thank L.~Dessart for applying the expanding photosphere method to
SN 2008bk to derive an approximate explosion date and A.~Howell for
access to his supernova spectra fit software \textsc{superfit}. The authors
gratefully acknowledge the support of the United States National
Science Foundation for the construction and operation of the LIGO
Laboratory, the Science and Technology Facilities Council of the
United Kingdom, the Max-Planck-Society and the State of
Niedersachsen/Germany for support of the construction and operation of
the GEO\,600 detector, and the Italian Istituto Nazionale di Fisica
Nucleare and the French Centre National de la Recherche Scientifique
for the construction and operation of the Virgo detector. The authors
also gratefully acknowledge the support of the research by these
agencies and by the Australian Research Council, the Council of
Scientific and Industrial Research of India, the Istituto Nazionale di
Fisica Nucleare of Italy, the Spanish Ministerio de Educaci\'on y
Ciencia, the Conselleria d'Economia Hisenda i Innovaci\'o of the
Govern de les Illes Balears, the Foundation for Fundamental Research
on Matter supported by the Netherlands Organisation for Scientific
Research, the Polish Ministry of Science and Higher Education, the
FOCUS Programme of Foundation for Polish Science, the Royal Society,
the Scottish Funding Council, the Scottish Universities Physics
Alliance, the National Aeronautics and Space Administration, the
Carnegie Trust, the Leverhulme Trust, the David and Lucile Packard
Foundation, the Research Corporation, and the Alfred P. Sloan
Foundation.
This document has been assigned LIGO Laboratory document number \ligodoc.

\end{acknowledgments}

\bibliographystyle{apsrev}

\end{document}